\documentclass[aps,prl,twocolumn,superscriptaddress,showpacs,nofootinbib,longbibliography]{revtex4-1}
\usepackage{graphicx}% Include figure files
\usepackage{dcolumn}% Align table columns on decimal point
\usepackage{bm}% bold math
\usepackage{hyperref}% add hypertext capabilities
\usepackage{upgreek}
\usepackage{color}
\usepackage{soul}
\usepackage{epstopdf}
\usepackage{units}
\usepackage{longtable}
\usepackage{floatrow}
\usepackage{latexsym}
\usepackage{gensymb}
\usepackage{floatrow}
\usepackage{mhchem}
\usepackage{xcolor}
\definecolor{mygreen}{RGB}{100,185,80}

\makeatletter
\renewcommand{\@fnsymbol}[1]{%
  \ensuremath{%
    \ifcase#1
      \or *            % 1
      \or \star        % 2 (statt †)
      \or \ddagger     % 3
      \or \mathsection % 4 (§)
      \or \diamond     % 5 (statt ¶/P)
      \or \|           % 6
    \else *\fi}}
\makeatother

\begin{document}

%I think we will need a more sexy title, a loose suggestion below
%\title{Epitaxial alignment without lattice matching in MBE grown superconducting FeSe films on SrTiO\textsubscript{3}}
\title{Unconventional incommensurate epitaxy of superconducting FeSe films on SrTiO\textsubscript{3}}

\author{M. Klement}
\email[email:]{martin.klement@uni-wuerzburg.de}
\affiliation{Julius-Maximilians-Universität Würzburg, Fakultät für Physik und Astronomie (EP3), Am Hubland, D-97074 Würzburg, Deutschland}
\affiliation{Julius-Maximilians-Universität Würzburg, Institut für Topologische Isolatoren, Am Hubland, D-97074 Würzburg, Deutschland}

%I put my name second, if someone objects we can of course change it
\author{K. M. Fijalkowski}
\email[email:]{kajetan.fijalkowski@uni-wuerzburg.de}
\affiliation{Julius-Maximilians-Universität Würzburg, Fakultät für Physik und Astronomie (EP3), Am Hubland, D-97074 Würzburg, Deutschland}
\affiliation{Julius-Maximilians-Universität Würzburg, Institut für Topologische Isolatoren, Am Hubland, D-97074 Würzburg, Deutschland}

\author{M. Kamp}
\email[email:]{martin.kamp@uni-wuerzburg.de}
\affiliation{Julius-Maximilians-Universität Würzburg, Physikalisches Institut, Am Hubland, D-97074 Würzburg, Deutschland}

\author{C. Gould}
\email[email:]{charles.gould@physik.uni-wuerzburg.de}
\affiliation{Julius-Maximilians-Universität Würzburg, Fakultät für Physik und Astronomie (EP3), Am Hubland, D-97074 Würzburg, Deutschland}
\affiliation{Julius-Maximilians-Universität Würzburg, Institut für Topologische Isolatoren, Am Hubland, D-97074 Würzburg, Deutschland}

\author{L. W. Molenkamp}
\email[email:]{molenkamp@physik.uni-wuerzburg.de}
\affiliation{Julius-Maximilians-Universität Würzburg, Fakultät für Physik und Astronomie (EP3), Am Hubland, D-97074 Würzburg, Deutschland}
\affiliation{Julius-Maximilians-Universität Würzburg, Institut für Topologische Isolatoren, Am Hubland, D-97074 Würzburg, Deutschland}

\date{\today}

\begin{abstract}
We present a combined X-ray diffraction and transmission electron microscopy study of superconducting FeSe/FeTe multilayers grown by molecular beam epitaxy on SrTiO\textsubscript{3}(001) substrates. While X-ray diffraction confirms perfect in-plane epitaxial alignment between FeSe, FeTe, and the substrate, scanning transmission electron microscopy reveals a surprising lack of atomic registry at the FeSe/SrTiO\textsubscript{3} interface. Instead of adapting to the substrate lattice, FeSe retains its own in-plane lattice spacing. A periodic lateral shift between the atomic positions of FeSe and SrTiO\textsubscript{3} is observed, with a registry recurrence length that matches the lattice mismatch determined by X-ray diffraction. No misfit dislocations or other relaxation features are detected at the interface. This coexistence of directional alignment and registry-free growth suggests an unconventional regime of epitaxy in which crystallographic orientation is maintained without atomic matching. The findings offer insight into strain accommodation in layered systems and may have implications for interface engineering in Fe-based superconductors.
\end{abstract}

\maketitle

\section[Introduction]{Introduction}

The iron chalcogenide FeSe is a currently topical material system in superconducting thin films. As one of the structurally simplest iron-based superconductors, bulk FeSe exhibits a superconducting transition temperature of 8 K \cite{hsu2008}. When grown as a monolayer on SrTiO\textsubscript{3}(001) via molecular beam epitaxy (MBE), transition temperature increases dramatically to values well above 40 K \cite{wang2012,ge2015}, possibly due to interfacial effects such as lattice strain \cite{peng2013_strain} and electron–phonon coupling \cite{lee2014_interfacial_mode_coupling}.

While the electronic properties of ultrathin FeSe have been extensively studied, structural aspects of growth—such as epitaxial alignment, lattice mismatch, and strain relaxation—remain less systematically explored. Previous scanning transmission electron microscopy (STEM) investigations on monolayer FeSe reveal lateral shifts at the FeSe/SrTiO\textsubscript{3} interface, suggesting registry mismatch and local structural distortions \cite{li2016_atomically,peng2020_picoscale}. While these works even propose possible atomic registries, they implicitly assume identical in-plane lattice constants and do not provide supporting evidence from X-ray diffraction (XRD).

This assumption is also widespread across the theoretical and modeling literature. Density functional theory (DFT) studies commonly fix the in-plane lattice constant of FeSe to that of the SrTiO\textsubscript{3} substrate ($a = 3.905$ \AA) \cite{yang2024_phonon, wang2016_epcoupling, cao2014_sdw}, even though this approximation has not been validated for FeSe thin films.
To our knowledge, no prior work has explicitly incorporated the experimentally observed lattice mismatch into structural models of thin-layer FeSe. This simplification is inconsistent with the high-resolution XRD and STEM results presented here, which demonstrate a relaxed, incommensurate interface without signs of pseudomorphic strain.

In this work, we investigate superconducting FeSe films with a thickness of ten monolayers grown on SrTiO\textsubscript{3}(001) by MBE. Using high-resolution XRD, azimuthal $\phi$-scans, and cross-sectional STEM, we analyze both global and local structural features to understand how the FeSe lattice adapts to the substrate.

Our STEM data reveal a periodic lateral shift between the atomic positions of FeSe and SrTiO\textsubscript{3}, with a repetition length matching the lattice mismatch determined by XRD. No misfit dislocations or plastic relaxation mechanisms are observed. This indicates a structurally continuous, orientation-coherent growth regime characterized by in-plane epitaxial alignment despite incommensurate, non-registered atomic positions — a mode that challenges the conventional notion of pseudomorphic epitaxy and provides new insights into interface engineering in layered superconductors.

\section[Introduction]{MBE Growth}

The FeSe thin films investigated in this study were grown on (001)-oriented SrTiO\textsubscript{3} single-crystal substrates purchased from Crystec GmbH. SrTiO\textsubscript{3} consists of alternating layers of SrO and TiO\textsubscript{2} along the [001] direction. To achieve atomically flat, TiO\textsubscript{2}-terminated surfaces, the substrates were chemically etched in buffered HF solution, which selectively removes the SrO termination layer, followed by annealing in a tube furnace at 980$^\circ$C for 3--4 hours in an oxygen atmosphere. This procedure is known to reproducibly yield TiO\textsubscript{2}-terminated step-terrace structures with single-unit-cell step heights \cite{gelle2018}. The presence of such unit-cell-high steps in our AFM measurements of the substrate (see Supplemental Material~\cite{SuppMat}) confirms the termination, as the selective removal of SrO during etching implies that the adjacent terraces necessarily expose TiO\textsubscript{2} surfaces.

Film deposition is performed in an ultra-high vacuum MBE system with a base pressure below (1 $\times$ 10$^{-10}$ mbar). The film consists of 10 monolayers of FeSe ($\sim$5.5 nm), followed by 33 monolayers of FeTe ($\sim$20.8 nm), and a thin amorphous Te capping layer ($\sim$4.5 nm). Prior to FeSe growth, the substrates are annealed in situ at 650$^\circ$C for 2 hours to desorb contaminants. The FeSe layer is deposited at 500$^\circ$C over 40 minutes with a Se:Fe beam equivalent pressure (BEP) ratio of approximately 10:1 to ensure selenium-rich conditions. A subsequent post-growth annealing at 700 $^\circ$C for 2 hours is applied to remove excess selenium from the film, as commonly reported in the literature \cite{Faeth2021, Liu2020, li2016_atomically}, and may also help to cure defects and reduce stacking faults.

Subsequently, an FeTe layer is deposited directly on top of the FeSe layer as a protective and electronically compatible cap. This choice is motivated by prior findings showing that crystalline Te capping can strongly suppress superconductivity in FeSe, whereas epitaxial FeTe maintains a high critical temperature with minimal degradation \cite{li2021_capping}. The top Te layer, deposited at low temperature, serves only as an oxidation barrier and does not contact the FeSe directly.

The FeTe layer is grown at 250$^\circ$C for 150 minutes, followed by Te deposition at a temperature between 0$^\circ$C and 10$^\circ$C for 2.5 minutes. All elements are evaporated from standard Knudsen cells. The Te:Fe beam equivalent pressure (BEP) ratio is set to $\sim$15:1 to ensure tellurium-rich conditions during FeTe growth. The entire growth process is monitored in situ via reflection of high energy electron diffraction (RHEED), allowing real-time assessment of growth mode transitions and surface ordering. Final film thicknesses are determined ex situ by cross-sectional STEM.

\section{Electrical transport and confirmation of superconductivity}

To confirm the superconducting properties of the FeSe/FeTe heterostructures, transport measurements are performed on samples that are nominally identical to those used for structural characterization. Due to the limited sample size after growth ($3\times5$ mm), XRD, STEM, and transport experiments are carried out on separate but equivalently prepared films. All samples are fabricated under identical MBE conditions and exhibit the same nominal thicknesses and growth parameters.

\begin{figure}[ht]
    \centering
\includegraphics[width=1\textwidth]{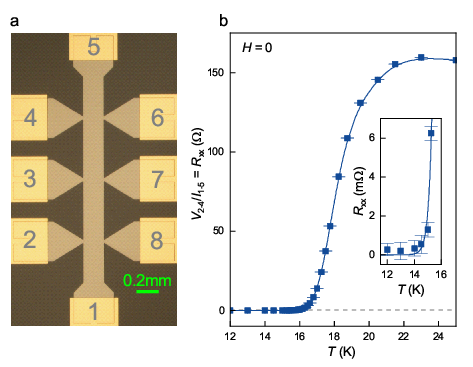}
    \caption{a: Optical micrograph of the eight-terminal device geometry used for resistance measurements. b: Four-terminal resistance as a function of temperature, showing a clear superconducting transition in a nominally identical FeSe/FeTe sample. Solid markers with error bars are the measured data; the smooth curve is only a guide to the eye to highlight the overall transition profile. The inset shows a zoomed-in segment of the data near zero resistance (note the units here are milliohms).}
    \label{fig:transport}
\end{figure}

For the transport measurements, samples are patterned into eight-terminal Hall bar geometries using a combination of optical and electron beam lithography. The distance between two adjacent side contacts is 600 $\mu$m, and the width of the Hall bar is 200 $\mu$m. Electrical contacts are made by bonding gold wires onto pre-defined Au pads. A photograph of a completed Hall bar device is shown in Figure \ref{fig:transport}a.

Resistance versus temperature measurements are performed using standard four-terminal low-frequency a.c. measurement techniques (with a bias current of 3 $\mu$A at 13.7 Hz), with the sample inside a liquid helium cryostat with a variable temperature. Figure \ref{fig:transport}b shows a representative $R$–$T$ curve, collected at zero external magnetic field, demonstrating a superconducting transition with a zero resistance state observed below some 14 K (see the inset to Figure \ref{fig:transport}b). 

The superconducting transition is markedly broadened when compared to more conventional BCS superconductors. In the usual BCS case, a sharp and clearly distinctive critical temperature can be discerned. The observed broadening is not surprising, as thin FeSe layers have previously been reported \cite{Schneider2014,Zhao2016,zhao2024_growth} to host the Berezinskii–Kosterlitz–Thouless (BKT) transition \cite{Kosterlitz1973,Halperin1979}, establishing them as 2D superconductors. The breakdown of superconductivity stems from proliferation of thermally activated superconducting vortices, that dissipate energy when moved by the electrical current. The higher the temperature, the more vortices are activated, which results in a resistance curve that smoothly lifts off from zero. Details of the BKT transition in our hetero-structures will be the subject of future work.

\section{X-ray Diffraction Analysis}

\begin{figure}[ht]
    \centering
    \includegraphics[width=1\textwidth]{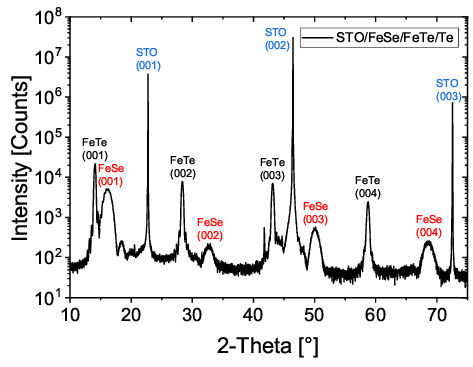}
    \caption{$\omega$-2$\theta$ scan of a FeSe thin film capped with FeTe, grown on a SrTiO\textsubscript{3}(001) substrate. Reflections from the substrate, FeSe, and FeTe are clearly visible. Pronounced Kiessig fringes appear as closely spaced oscillations around the FeTe reflections. Weaker and more widely spaced Kiessig oscillations can also be discerned around the FeSe(001) reflection at approximately $18.5^\circ$, consistent with the thinner FeSe layer. A high-resolution view of the FeTe(001) reflection, highlighting the surrounding Kiessig oscillations, is shown in Fig.~3, where the Kiessig oscillations are indicated by arrows.}
    \label{fig:XRD_sym}
\end{figure}

XRD is used to analyze the structural properties of the FeSe/FeTe multilayer. The $\omega$–2$\theta$ scan in Figure \ref{fig:XRD_sym} reveals well-defined (00$l$) reflections from the SrTiO\textsubscript{3} substrate, the FeSe base layer, and the overlying FeTe film. Both layers exhibit clear interference fringes, indicating sharp interfaces and uniform thicknesses. In particular, multiple Kiessig fringes—interference oscillations arising from reflections at the film surfaces and interfaces \cite{miller2022_laue_oscillations}—are visible around the FeTe(001) and (003) peaks, while weaker oscillations are also discernible at the FeSe(001) and (002) reflections.

Out-of-plane lattice constants are extracted from several (00$l$) reflections of both FeSe and FeTe using the Bragg equation. The peak positions are referenced to the strong SrTiO\textsubscript{3}(002) substrate signal, which serves both for instrument alignment and as an internal standard for determining the FeSe and FeTe peak positions. This yields $c$-axis lattice parameters of 5.468 \AA{} for FeSe and 6.284 \AA{} for FeTe, with uncertainties on the order of 0.01 \AA.

\begin{figure}[htb]
  \centering
  \includegraphics[width=1\linewidth]{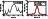}
  \caption{a: Kiessig fringes at the FeTe(001) reflection (indicated by arrows) used to extract a thickness of $20.8\,\mathrm{nm}$. b: Rocking curve of the FeSe(003) peak showing a FWHM of $0.036^\circ$ (130 arcsec), indicating exceptional crystalline quality and alignment.
  }
  \label{fig:XRD_kiessig_rocking}
\end{figure}

From the Kiessig fringe spacing in Figure \ref{fig:XRD_kiessig_rocking}a, the FeTe layer thickness is determined to be 20.8 nm, in excellent agreement with the nominal thickness of 33 monolayers ($\sim$20.73 nm). For FeSe, a single fringe near the (002) reflection suggests a thickness of approximately 4.2 nm, consistent with the expected value of 10 monolayers ($\sim$5.47 nm), albeit with increased uncertainty due to the weaker signal, measurement noise, and the fact that only one fringe is visible—unlike the multiple oscillations observed for FeTe (see Supplemental Material~\cite{SuppMat}).

Crystalline quality of the FeSe layer is further evaluated via the (003) rocking curve in Figure \ref{fig:XRD_kiessig_rocking}b, which yields a full width at half maximum (FWHM) of 0.036° (130 arcsec). Previous reports on FeSe films on SrTiO\textsubscript{3} and CaF\textsubscript{2} substrates show FWHM values ranging from approximately 0.12° to 0.61° \cite{zhao2024_growth, huang2021_directARPES, feng2018_tunable}. The exceptionally narrow rocking curve underscores the outstanding crystalline coherence of our MBE-grown films.

To determine the in-plane lattice parameters, asymmetrical (112) reflections are measured in a tilted geometry (see Supplemental Material~\cite{SuppMat}). Based on the known tetragonal symmetry of FeSe and FeTe, the in-plane lattice constants of our films are determined to be 3.862 \AA{} for FeSe and 3.824 \AA{} for FeTe. The FeSe layer therefore exhibits an in-plane lattice mismatch of approximately $-1.1\%$ (compared to the SrTiO$_3$ substrate ($a_{\mathrm{STO}} = 3.905$ \AA)). In contrast, bulk FeSe has a smaller lattice parameter of about 3.765 \AA{}  \cite{hsu2008}, resulting in an apparent expansion in the thin-film regime. This observation is consistent with previous reports showing that ultra-thin FeSe films can exhibit an increased in-plane lattice constant compared to the bulk, particularly for one-unit-cell FeSe on SrTiO\textsubscript{3} \cite{liu2020_fe_se_2dmaterials}. Various mechanisms have been proposed to explain this effect, including tensile strain \cite{peng2014_tuning} and interface-induced modifications of the electronic structure and spin-ordering phenomena \cite{tan2013_interface}.

\begin{figure}[ht]
    \centering
\includegraphics[width=1\textwidth]{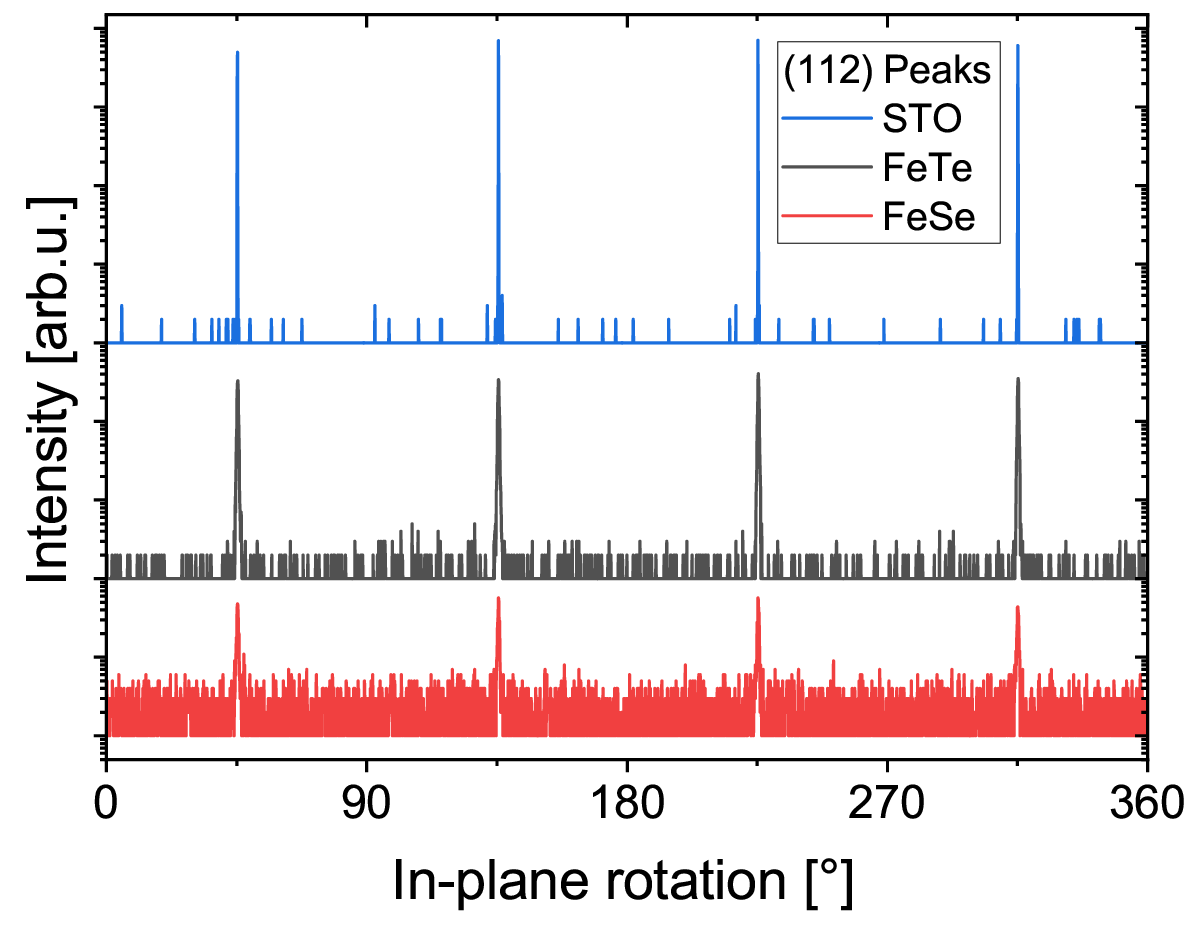}
    \caption{Azimuthal $\phi$-scans of the (112) reflections from SrTiO\textsubscript{3} (blue), FeSe (red), and FeTe (black). All three scans exhibit a clear fourfold symmetry with peaks spaced by 90°, indicating a shared in-plane crystallographic orientation. The peak positions coincide at $\phi = 45^\circ$, $135^\circ$, $225^\circ$, and $315^\circ$, confirming an untwinned, orientation-coherent in-plane epitaxial relationship between the film layers and the substrate. Vertical offsets have been applied for clarity.}
    \label{fig:XRD_PhiScan}
\end{figure}

The azimuthal $\phi$-scans of the (112) reflections from FeSe, FeTe, and SrTiO\textsubscript{3}, shown in Figure \ref{fig:XRD_PhiScan}, exhibit a clear fourfold symmetry with all reflections aligned at $\phi = 45^\circ$ and every subsequent 90°. This indicates an orientation-coherent in-plane epitaxial relationship between all three materials, with no observable rotation or twinning. Despite the lattice mismatch discussed above, the in-plane crystallographic axes of FeSe and FeTe are aligned with those of the substrate. This epitaxial registry is a key observation, especially in light of the relaxation behavior discussed in the following STEM section.

\section{Transmission Electron Microscopy Analysis}

\begin{figure}[ht]
  \centering
  \includegraphics[width=\linewidth]{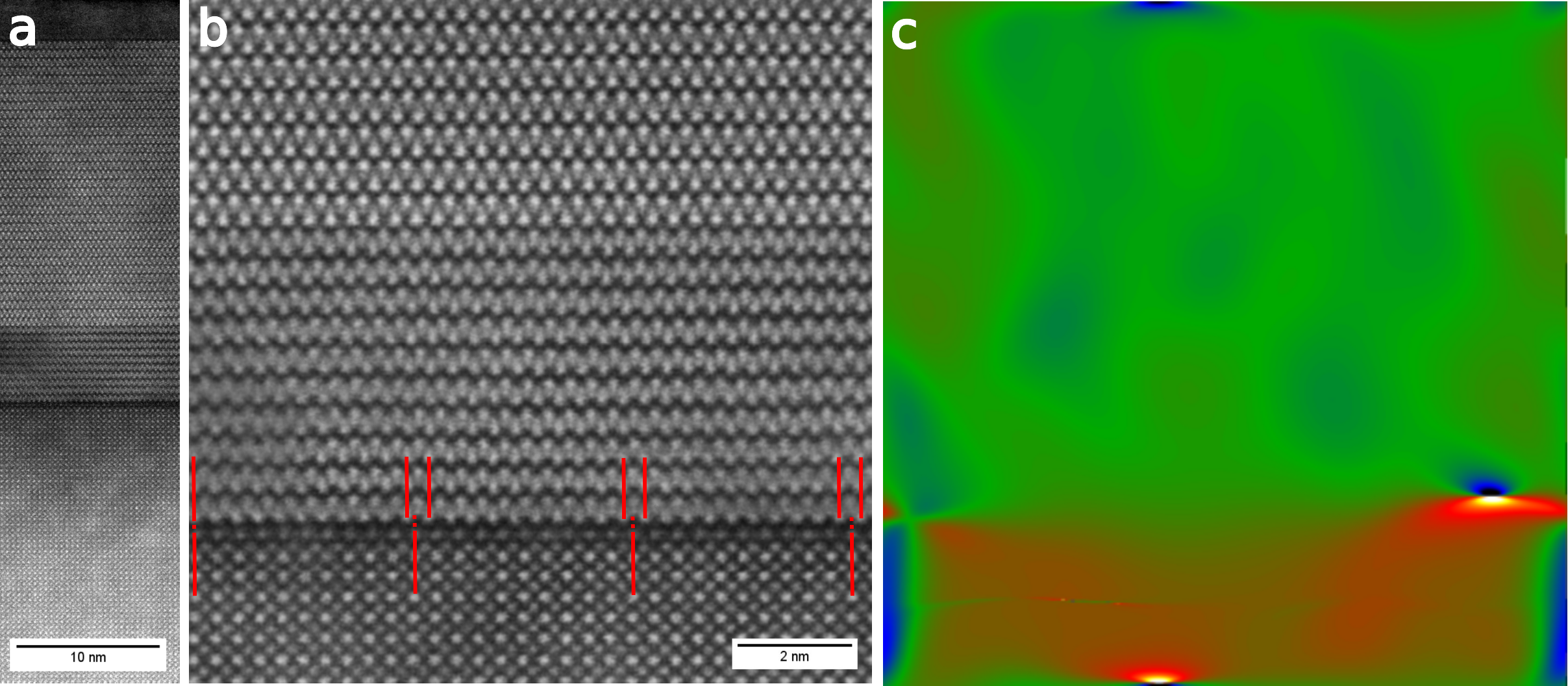}
  \caption{High-resolution HAADF-STEM images of the FeSe/SrTiO$_3$ heterostructure. 
(a) Overview of the full stack, showing the SrTiO$_3$ substrate, a TiO$_2$ double layer, epitaxial FeSe and FeTe films, and the amorphous tellurium capping layer. 
(b) Magnified view of the FeSe/SrTiO$_3$ interface. Vertical red lines mark selected unit cell positions in both SrTiO$_3$ and FeSe along the in-plane direction. A progressive lateral shift of the FeSe lattice is observed, reaching a half-unit-cell offset after approximately 30 unit cells and full registry after approximately 60. The TiO$_2$ double layer is clearly resolved at the interface. 
(c) Geometric phase analysis (GPA) map derived from panel (b), showing the spatial distribution of relative lattice-parameter variations. A clear contrast between substrate and film is observed, reflecting their different lattice parameters, while the film region exhibits a spatially uniform distribution without localized distortions within the experimental resolution.}
  \label{fig:tem-overview}
\end{figure}

We now turn to cross-sectional high-angle annular dark-field scanning transmission electron microscopy (HAADF-STEM) to assess the structural quality of the FeSe/FeTe multilayers and their interfaces with the SrTiO\textsubscript{3}(001) substrate, as shown in Figure \ref{fig:tem-overview}. 

Panel a presents an overview of the full heterostructure stack, comprising the SrTiO\textsubscript{3} substrate, a distinct interfacial layer, the epitaxial FeSe and FeTe films, and the amorphous tellurium capping layer. The interfacial layer between SrTiO\textsubscript{3} and FeSe exhibits a characteristic darker contrast in STEM images. Based on its thickness and morphology, and in agreement with previous experimental and theoretical studies \cite{erdman2002_tio2, herger2007_srtio3, zhu2012_reconstruction, kienzle2011_vacant, marshall2015_defects, li2016_atomically}, this layer is attributed to a TiO\textsubscript{2} double-layer termination of the SrTiO\textsubscript{3}(001) surface. The overall structure is continuous and uniform across the field of view, with no evidence of amorphous regions, delamination, or interfacial disruption. All interfaces are atomically abrupt and structurally continuous, with no indication of defects or misfit accommodation. Across the 8 $\mu$m-long STEM lamella, approximately 50 regions were imaged at various magnifications, including atomic-resolution STEM. The surveyed area covers approximately 25\% of the total lamella length. In none of these regions were interface defects, lattice mismatch accommodation, or any structural irregularities at the heterointerfaces observed.

Panel b is a magnified view of the FeSe/SrTiO$_3$ interface with overlaid 
markers indicating selected unit cell positions in both materials. 
The lateral atomic alignment between FeSe and SrTiO$_3$ is initially well aligned 
on the left side of the image, but a systematic in-plane shift becomes 
apparent across the field of view. After approximately 30 unit cells, 
the FeSe lattice is laterally displaced by half a unit cell relative to 
the substrate; full registry recurs after approximately 60 unit cells. 
Using SrTiO$_3$ with an in-plane lattice constant of 3.905 \AA{} as a reference, this corresponds to an in-plane lattice constant of approximately 3.841 \AA{} for FeSe, in excellent agreement with the value obtained from a numerical one-dimensional Fourier amplitude analysis of multiple STEM images (see Supplemental Material~\cite{SuppMat}, Fig.~S8), where a weighted mean of $a_{\mathrm{FeSe}} = 3.843$ \AA{} is extracted, as well as with the value extracted from X-ray diffraction.

This gradual mismatch accumulation is accommodated continuously across the film thickness without the formation of misfit dislocations or other extended defects. The lattice planes remain intact across the entire transition region, indicating that the structural adjustment occurs smoothly rather than through discrete defect formation.

To examine possible local lattice-parameter variations, a geometric phase analysis (GPA) was performed on the high-resolution STEM image shown in Fig.~\ref{fig:tem-overview}b. The corresponding map is presented in Fig.~\ref{fig:tem-overview}c. A clear contrast between the SrTiO$_3$ substrate and the FeSe/FeTe film is observed, reflecting their different lattice parameters. Within the film region, however, the distribution is spatially uniform and does not exhibit localized variations or gradients within the experimental resolution. In particular, no distinct contrast is detected at the FeSe/FeTe interface.

Taken together with the XRD results, these observations indicate that the film retains its own in-plane lattice spacing and is fully relaxed with respect to the SrTiO$_3$ substrate within the limits of the experimental resolution.

\begin{figure}[ht]
  \centering
  \includegraphics[width=\linewidth]{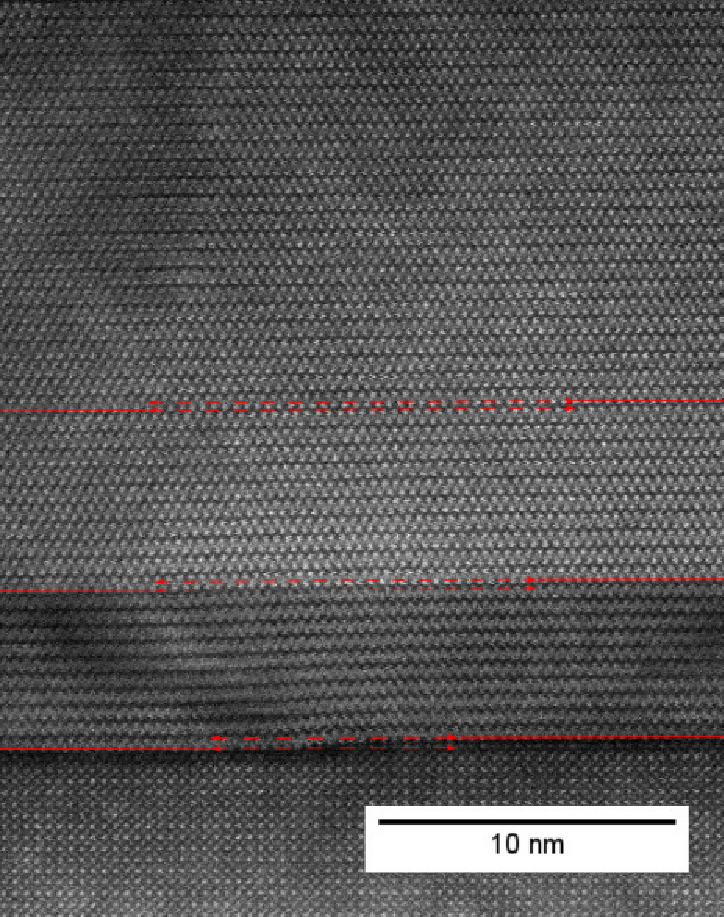}
  \caption{HAADF-STEM image of a monoatomic step at the SrTiO\textsubscript{3} surface and the resulting lateral displacement in the film. Solid red lines mark corresponding atomic planes on either side of the step as long as the atomic positions remain aligned. Where the lattice begins to deviate, dashed red lines are used to continue the trajectories and highlight the height difference across the step. The transition becomes progressively broader and flatter with increasing distance from the substrate, from the first FeSe monolayers into the upper FeTe layers.}
  \label{fig:TEM_2}
\end{figure}

Due to the (unintentional) miscut of the substrate, atomic terraces and monoatomic step edges are present at the surface. In one region of the STEM lamella, such a step is captured, as shown in Fig.~\ref{fig:TEM_2}. The FeSe layer continues laterally uninterrupted over the step, but instead of reproducing the sharp contour of the underlying terrace edge, it exhibits a smooth transition. The film effectively bridges the step with a gentle curvature, resembling the way a flexible membrane conforms over a discontinuity. This behavior indicates a two-dimensional growth mode that allows uniform lateral coverage across minor variations in surface topography, without introducing morphological discontinuities.

Solid red lines mark corresponding atomic planes on either side of the step, extending as long as the atom positions remain aligned. Where the lattice begins to deviate, dashed red lines are used to continue the trajectory and highlight the height difference across the step. These vertical and lateral offsets become progressively broader and shallower, respectively, with increasing distance from the substrate—from the first monolayer of FeSe to the upper FeTe layers—reflecting a gradual reduction of the local lattice distortion associated with the substrate step.

Based on the substrate miscut of approximately 0.1°, corresponding to one unit-cell-high step approximately every 200~nm, only a small geometrical fraction of the film lies in the immediate vicinity of such step edges, where lattice-plane bending is visible in the STEM images.

Importantly, despite the presence of this atomic step, the crystallinity of the film remains intact across the entire transition region. High-resolution STEM images reveal no signs of dislocations, stacking faults, or other crystallographic defects, underscoring the high epitaxial quality and coherence of the heterostructure even in the presence of local topographical variations.

\section{Discussion}

Building on our high-resolution STEM analysis, we observe an intriguing coexistence of perfect in-plane alignment and persistent atomic misregistry at the FeSe/SrTiO\textsubscript{3} interface.
This structural configuration is highly unusual and not commonly addressed in the literature on FeSe heterostructures. Although FeSe and SrTiO\textsubscript{3} exhibit a lattice mismatch of approximately 1.1\%, the interface nevertheless remains defect-free: FeSe retains its characteristic thin-film in-plane lattice constant while maintaining directional alignment with the substrate. In typical lattice-mismatched systems, such discrepancies are compensated by strain accumulation, misfit dislocations, or partial epitaxial relaxation. However, none of these mechanisms are observed here.

This observation challenges the conventional understanding of epitaxy in lattice-mismatched systems. It suggests that directional alignment can be preserved even without atomic registry—possibly due to interfacial energy minimization, weak interlayer bonding, or subtle reconstruction effects at the atomic scale. Remarkably, the STEM data show no evidence of misfit dislocations, stacking faults, or other structural defects, further highlighting the high crystalline quality and the ability of the layered system to accommodate mismatch without introducing local distortions.

A central question that arises is how registry-free epitaxy can be stabilized under these conditions. TiO\textsubscript{2}-terminated SrTiO\textsubscript{3}(001) surfaces are known to exhibit under-coordinated Ti atoms, oxygen vacancies, and local structural relaxations \cite{zhu2012_reconstruction, jeschke2015oxygen, erdman2002_tio2}. These structural features can facilitate flexible electronic interactions and weak, adaptable bonding at the interface. Our high-resolution STEM images (Figs. \ref{fig:tem-overview} and \ref{fig:TEM_2}) clearly resolve the sharp and atomically abrupt interfaces between SrTiO\textsubscript{3}, a distinct TiO\textsubscript{2} interfacial double layer, and the FeSe film. The absence of any visible large-scale periodic surface reconstruction patterns indicates that significant reconstruction effects are not present at the FeSe/STO interface. However, subtle local structural relaxations or minor atomic displacements cannot be excluded based solely on the present STEM data. Such local adjustments could enable epitaxial alignment without introducing misfit dislocations or significant strain, even in the presence of lattice mismatch. 

We also note, that this growth mode exhibits different
features than conventional van der Waals epitaxy \cite{Koma1992,Walsh2017}. 
The atomic interfaces we observe are sharp and defect-free, 
while maintaining excellent structural continuity over a large distance, 
without any visible domain structure, all of which are hurdles 
often encountered for conventional van der Waals epitaxial layers. 
Indeed, in that regard, our layers are much more akin to conventional 
hetero-epitaxial films, with stronger, ionic bonding between the epilayer 
and the substrate. 
How that is possible without atomic registry or any large-scale interface 
reconstruction should be an interesting extension of this research.

This behavior does not fit any of the classical epitaxial categories. 
The FeSe layers are neither coherently strained 
($a_{\mathrm{FeSe}} = 3.862$ \AA{} rather than adopting $a_{\mathrm{SrTiO_3}} = 3.905 $ \AA) 
nor semi-coherent (which would involve the formation of misfit dislocations), 
yet they maintain perfect in-plane epitaxial alignment. 
This combination of in-plane incommensurability, dislocation-free interfaces, 
and stable azimuthal orientation represents an unconventional epitaxial regime 
that is not captured by standard strain–relaxation models.

To fully understand the origin of this registry-free epitaxy, additional investigations are needed. In-situ growth monitoring, atomistic interface simulations, or local spectroscopy may provide deeper insight into this behavior and clarify whether similar registry-free epitaxial modes occur in other hetero-structures.

\section{Conclusions}

We investigate the structural properties of MBE-grown superconducting FeSe/FeTe multilayers on SrTiO\textsubscript{3}(001) substrates using a combination of high-resolution XRD and STEM analysis. The films exhibit outstanding crystalline quality with atomically sharp interfaces and continuous, defect-free overgrowth of surface steps.

XRD measurements confirm that both FeSe and FeTe grow epitaxially aligned with the substrate while preserving their intrinsic lattice constants. A periodic lateral offset between FeSe and the substrate, consistent with the lattice mismatch, is observed and accommodated without the formation of dislocations.

STEM analysis reveals that atomic terraces and step edges are bridged smoothly by the film. The resulting transition zone spans several nanometers and is distributed over a larger lateral distance in upper layers, reflecting the film’s ability to gradually relax the vertical offset geometrically rather than plastically. The overall registry-free epitaxy, absence of defects, and preservation of crystallographic orientation underscore an unconventional but effective mode of lattice accommodation. The resulting ordering is reminiscent of a line Moir\'e pattern with a Moir\'e wavelength in the order of 30 nm developing between two materials having different lattice constants.

These findings not only demonstrate the mechanical adaptability of FeSe/FeTe heterostructures but also raise fundamental questions about the role of interfacial interactions in determining epitaxial behavior. Further investigations, including theoretical modeling and atomically resolved spectroscopy, are warranted to better understand the implications of this unique epitaxial mode for electronic structure and superconductivity.

\section[Data Availability]{Data Availability}
The data that support the findings of this study are openly available in the Zenodo repository at \url{https://doi.org/10.5281/zenodo.20050166}.

\section[Conflict of Interest]{Conflict of Interest}
The authors declare no conflict of interest.

\section[Author Contribution]{Author Contribution}
M. Kl. prepared the substrates, carried out the MBE growth of the samples, performed the XRD measurements, and analyzed the structural data as well as the STEM images. K. M. F. fabricated the lithographic devices and performed the transport measurements. M. Ka. prepared the STEM lamella and performed the imaging. C. G. and L. W. M. supervised the work. All authors contributed to the interpretation of the results and to the writing of the manuscript.

\section[Acknowledgents]{Acknowledgents}
We thank M. Stanley (Penn State) for in-depth discussions on growth parameters, flux calibration, and control of interface quality.
We gratefully acknowledge the financial support of the Free State of Bavaria (the Institute for Topological Insulators), Deutsche Forschungsgemeinschaft (SFB 1170, 258499086), W\"urzburg-Dresden Cluster of Excellence on Complexity, Topology and Dynamics in Quantum Matter (EXC 2147, 390858490), the European Commission under the H2020 FETPROACT Grant TOCHA (824140).

\section[References]{References}

%\bibliographystyle{unsrt}
%\bibliography{references}

\begin{thebibliography}{35}%
\makeatletter
\providecommand \@ifxundefined [1]{%
 \@ifx{#1\undefined}
}%
\providecommand \@ifnum [1]{%
 \ifnum #1\expandafter \@firstoftwo
 \else \expandafter \@secondoftwo
 \fi
}%
\providecommand \@ifx [1]{%
 \ifx #1\expandafter \@firstoftwo
 \else \expandafter \@secondoftwo
 \fi
}%
\providecommand \natexlab [1]{#1}%
\providecommand \enquote  [1]{``#1''}%
\providecommand \bibnamefont  [1]{#1}%
\providecommand \bibfnamefont [1]{#1}%
\providecommand \citenamefont [1]{#1}%
\providecommand \href@noop [0]{\@secondoftwo}%
\providecommand \href [0]{\begingroup \@sanitize@url \@href}%
\providecommand \@href[1]{\@@startlink{#1}\@@href}%
\providecommand \@@href[1]{\endgroup#1\@@endlink}%
\providecommand \@sanitize@url [0]{\catcode `\\12\catcode `\$12\catcode `\&12\catcode `\#12\catcode `\^12\catcode `\_12\catcode `\%12\relax}%
\providecommand \@@startlink[1]{}%
\providecommand \@@endlink[0]{}%
\providecommand \url  [0]{\begingroup\@sanitize@url \@url }%
\providecommand \@url [1]{\endgroup\@href {#1}{\urlprefix }}%
\providecommand \urlprefix  [0]{URL }%
\providecommand \Eprint [0]{\href }%
\providecommand \doibase [0]{http://dx.doi.org/}%
\providecommand \selectlanguage [0]{\@gobble}%
\providecommand \bibinfo  [0]{\@secondoftwo}%
\providecommand \bibfield  [0]{\@secondoftwo}%
\providecommand \translation [1]{[#1]}%
\providecommand \BibitemOpen [0]{}%
\providecommand \bibitemStop [0]{}%
\providecommand \bibitemNoStop [0]{.\EOS\space}%
\providecommand \EOS [0]{\spacefactor3000\relax}%
\providecommand \BibitemShut  [1]{\csname bibitem#1\endcsname}%
\let\auto@bib@innerbib\@empty
%</preamble>
\bibitem [{\citenamefont {Hsu}\ \emph {et~al.}(2008)\citenamefont {Hsu}, \citenamefont {Luo}, \citenamefont {Yeh}, \citenamefont {Chen}, \citenamefont {Huang}, \citenamefont {Wu}, \citenamefont {Lee}, \citenamefont {Huang}, \citenamefont {Chu}, \citenamefont {Yan},\ and\ \citenamefont {Wu}}]{hsu2008}%
  \BibitemOpen
  \bibfield  {author} {\bibinfo {author} {\bibfnamefont {F.-C.}\ \bibnamefont {Hsu}}, \bibinfo {author} {\bibfnamefont {J.-Y.}\ \bibnamefont {Luo}}, \bibinfo {author} {\bibfnamefont {K.-W.}\ \bibnamefont {Yeh}}, \bibinfo {author} {\bibfnamefont {T.-K.}\ \bibnamefont {Chen}}, \bibinfo {author} {\bibfnamefont {T.-W.}\ \bibnamefont {Huang}}, \bibinfo {author} {\bibfnamefont {P.~M.}\ \bibnamefont {Wu}}, \bibinfo {author} {\bibfnamefont {Y.-C.}\ \bibnamefont {Lee}}, \bibinfo {author} {\bibfnamefont {Y.-L.}\ \bibnamefont {Huang}}, \bibinfo {author} {\bibfnamefont {Y.-Y.}\ \bibnamefont {Chu}}, \bibinfo {author} {\bibfnamefont {D.-C.}\ \bibnamefont {Yan}}, \ and\ \bibinfo {author} {\bibfnamefont {M.-K.}\ \bibnamefont {Wu}},\ }\bibfield  {title} {\enquote {\bibinfo {title} {Superconductivity in the pbo-type structure $\alpha$-fese},}\ }\href {\doibase 10.1073/pnas.0807325105} {\bibfield  {journal} {\bibinfo  {journal} {Proc. Natl. Acad. Sci. USA}\ }\textbf {\bibinfo {volume} {105}},\ \bibinfo {pages} {14262--14264}
  (\bibinfo {year} {2008})}\BibitemShut {NoStop}%
\bibitem [{\citenamefont {Wang}\ \emph {et~al.}(2012)\citenamefont {Wang}, \citenamefont {Li}, \citenamefont {Zhang}, \citenamefont {Zhang}, \citenamefont {Zhang}, \citenamefont {Li}, \citenamefont {Ding}, \citenamefont {Ou}, \citenamefont {Deng}, \citenamefont {Chang}, \citenamefont {Wang}, \citenamefont {Ma}, \citenamefont {Chen}, \citenamefont {Xu}, \citenamefont {Chen}, \citenamefont {Zhou},\ and\ \citenamefont {Xue}}]{wang2012}%
  \BibitemOpen
  \bibfield  {author} {\bibinfo {author} {\bibfnamefont {Q.-Y.}\ \bibnamefont {Wang}}, \bibinfo {author} {\bibfnamefont {Z.}~\bibnamefont {Li}}, \bibinfo {author} {\bibfnamefont {W.-H.}\ \bibnamefont {Zhang}}, \bibinfo {author} {\bibfnamefont {Z.-C.}\ \bibnamefont {Zhang}}, \bibinfo {author} {\bibfnamefont {J.-S.}\ \bibnamefont {Zhang}}, \bibinfo {author} {\bibfnamefont {W.}~\bibnamefont {Li}}, \bibinfo {author} {\bibfnamefont {H.}~\bibnamefont {Ding}}, \bibinfo {author} {\bibfnamefont {Y.-B.}\ \bibnamefont {Ou}}, \bibinfo {author} {\bibfnamefont {P.}~\bibnamefont {Deng}}, \bibinfo {author} {\bibfnamefont {K.}~\bibnamefont {Chang}}, \bibinfo {author} {\bibfnamefont {J.}~\bibnamefont {Wang}}, \bibinfo {author} {\bibfnamefont {X.-C.}\ \bibnamefont {Ma}}, \bibinfo {author} {\bibfnamefont {X.}~\bibnamefont {Chen}}, \bibinfo {author} {\bibfnamefont {Z.-A.}\ \bibnamefont {Xu}}, \bibinfo {author} {\bibfnamefont {C.}~\bibnamefont {Chen}}, \bibinfo {author} {\bibfnamefont {X.~J.}\ \bibnamefont {Zhou}}, \ and\ \bibinfo
  {author} {\bibfnamefont {Q.-K.}\ \bibnamefont {Xue}},\ }\bibfield  {title} {\enquote {\bibinfo {title} {Interface-induced high-temperature superconductivity in single unit-cell fese films on srtio\textsubscript{3}},}\ }\href {\doibase 10.1088/0256-307X/29/3/037402} {\bibfield  {journal} {\bibinfo  {journal} {Chin. Phys. Lett.}\ }\textbf {\bibinfo {volume} {29}},\ \bibinfo {pages} {037402} (\bibinfo {year} {2012})}\BibitemShut {NoStop}%
\bibitem [{\citenamefont {Ge}\ \emph {et~al.}(2015)\citenamefont {Ge}, \citenamefont {Liu}, \citenamefont {Liu}, \citenamefont {Gao}, \citenamefont {Qian}, \citenamefont {Xue}, \citenamefont {Liu},\ and\ \citenamefont {Jia}}]{ge2015}%
  \BibitemOpen
  \bibfield  {author} {\bibinfo {author} {\bibfnamefont {J.-F.}\ \bibnamefont {Ge}}, \bibinfo {author} {\bibfnamefont {Z.-L.}\ \bibnamefont {Liu}}, \bibinfo {author} {\bibfnamefont {C.}~\bibnamefont {Liu}}, \bibinfo {author} {\bibfnamefont {C.-L.}\ \bibnamefont {Gao}}, \bibinfo {author} {\bibfnamefont {D.}~\bibnamefont {Qian}}, \bibinfo {author} {\bibfnamefont {Q.-K.}\ \bibnamefont {Xue}}, \bibinfo {author} {\bibfnamefont {Y.}~\bibnamefont {Liu}}, \ and\ \bibinfo {author} {\bibfnamefont {J.-F.}\ \bibnamefont {Jia}},\ }\bibfield  {title} {\enquote {\bibinfo {title} {Superconductivity above 100 k in single-layer fese films on doped srtio\textsubscript{3}},}\ }\href {\doibase 10.1038/nmat4153} {\bibfield  {journal} {\bibinfo  {journal} {Nat. Mater.}\ }\textbf {\bibinfo {volume} {14}},\ \bibinfo {pages} {285--289} (\bibinfo {year} {2015})}\BibitemShut {NoStop}%
\bibitem [{\citenamefont {Peng}\ \emph {et~al.}(2014{\natexlab{a}})\citenamefont {Peng}, \citenamefont {Xu}, \citenamefont {Tan}, \citenamefont {Cao}, \citenamefont {Xia}, \citenamefont {Shen}, \citenamefont {Huang}, \citenamefont {Wen}, \citenamefont {Song}, \citenamefont {Zhang}, \citenamefont {Xie}, \citenamefont {Hu},\ and\ \citenamefont {Feng}}]{peng2013_strain}%
  \BibitemOpen
  \bibfield  {author} {\bibinfo {author} {\bibfnamefont {R.}~\bibnamefont {Peng}}, \bibinfo {author} {\bibfnamefont {H.~C.}\ \bibnamefont {Xu}}, \bibinfo {author} {\bibfnamefont {S.~Y.}\ \bibnamefont {Tan}}, \bibinfo {author} {\bibfnamefont {H.~Y.}\ \bibnamefont {Cao}}, \bibinfo {author} {\bibfnamefont {M.}~\bibnamefont {Xia}}, \bibinfo {author} {\bibfnamefont {X.~P.}\ \bibnamefont {Shen}}, \bibinfo {author} {\bibfnamefont {Z.~C.}\ \bibnamefont {Huang}}, \bibinfo {author} {\bibfnamefont {C.~H.~P.}\ \bibnamefont {Wen}}, \bibinfo {author} {\bibfnamefont {Q.}~\bibnamefont {Song}}, \bibinfo {author} {\bibfnamefont {T.}~\bibnamefont {Zhang}}, \bibinfo {author} {\bibfnamefont {B.~P.}\ \bibnamefont {Xie}}, \bibinfo {author} {\bibfnamefont {X.~G.}\ \bibnamefont {Hu}}, \ and\ \bibinfo {author} {\bibfnamefont {D.~L.}\ \bibnamefont {Feng}},\ }\bibfield  {title} {\enquote {\bibinfo {title} {Measurement of an enhanced superconducting phase and a pronounced anisotropy of the energy gap of a strained {FeSe} single layer in
  fese/nbsrtio$_3$/ktao$_3$ heterostructures using photoemission spectroscopy},}\ }\href {\doibase 10.1103/PhysRevLett.112.107001} {\bibfield  {journal} {\bibinfo  {journal} {Physical Review Letters}\ }\textbf {\bibinfo {volume} {112}},\ \bibinfo {pages} {107001} (\bibinfo {year} {2014}{\natexlab{a}})}\BibitemShut {NoStop}%
\bibitem [{\citenamefont {Lee}\ \emph {et~al.}(2014)\citenamefont {Lee}, \citenamefont {Schmitt}, \citenamefont {Moore}, \citenamefont {Johnston}, \citenamefont {Cui}, \citenamefont {Li}, \citenamefont {Yi}, \citenamefont {Liu}, \citenamefont {Hashimoto}, \citenamefont {Zhang}, \citenamefont {Lu}, \citenamefont {Devereaux}, \citenamefont {Lee},\ and\ \citenamefont {Shen}}]{lee2014_interfacial_mode_coupling}%
  \BibitemOpen
  \bibfield  {author} {\bibinfo {author} {\bibfnamefont {J.~J.}\ \bibnamefont {Lee}}, \bibinfo {author} {\bibfnamefont {F.~T.}\ \bibnamefont {Schmitt}}, \bibinfo {author} {\bibfnamefont {R.~G.}\ \bibnamefont {Moore}}, \bibinfo {author} {\bibfnamefont {S.}~\bibnamefont {Johnston}}, \bibinfo {author} {\bibfnamefont {Y.-T.}\ \bibnamefont {Cui}}, \bibinfo {author} {\bibfnamefont {W.}~\bibnamefont {Li}}, \bibinfo {author} {\bibfnamefont {M.}~\bibnamefont {Yi}}, \bibinfo {author} {\bibfnamefont {Z.~K.}\ \bibnamefont {Liu}}, \bibinfo {author} {\bibfnamefont {M.}~\bibnamefont {Hashimoto}}, \bibinfo {author} {\bibfnamefont {Y.}~\bibnamefont {Zhang}}, \bibinfo {author} {\bibfnamefont {D.~H.}\ \bibnamefont {Lu}}, \bibinfo {author} {\bibfnamefont {T.~P.}\ \bibnamefont {Devereaux}}, \bibinfo {author} {\bibfnamefont {D.-H.}\ \bibnamefont {Lee}}, \ and\ \bibinfo {author} {\bibfnamefont {Z.-X.}\ \bibnamefont {Shen}},\ }\bibfield  {title} {\enquote {\bibinfo {title} {Interfacial mode coupling as the origin of the enhancement
  of $t_c$ in fese films on srtio$_3$},}\ }\href {\doibase 10.1038/nature13894} {\bibfield  {journal} {\bibinfo  {journal} {Nature}\ }\textbf {\bibinfo {volume} {515}},\ \bibinfo {pages} {245--248} (\bibinfo {year} {2014})}\BibitemShut {NoStop}%
\bibitem [{\citenamefont {Li}\ \emph {et~al.}(2016)\citenamefont {Li}, \citenamefont {Zhang}, \citenamefont {Tang}, \citenamefont {Liu}, \citenamefont {Shi}, \citenamefont {Nie}, \citenamefont {Zhou}, \citenamefont {Li}, \citenamefont {Zhang}, \citenamefont {Song}, \citenamefont {He}, \citenamefont {Ji}, \citenamefont {Zhang}, \citenamefont {Gu}, \citenamefont {Wang}, \citenamefont {Ma},\ and\ \citenamefont {Xue}}]{li2016_atomically}%
  \BibitemOpen
  \bibfield  {author} {\bibinfo {author} {\bibfnamefont {F.}~\bibnamefont {Li}}, \bibinfo {author} {\bibfnamefont {Q.}~\bibnamefont {Zhang}}, \bibinfo {author} {\bibfnamefont {C.}~\bibnamefont {Tang}}, \bibinfo {author} {\bibfnamefont {C.}~\bibnamefont {Liu}}, \bibinfo {author} {\bibfnamefont {J.}~\bibnamefont {Shi}}, \bibinfo {author} {\bibfnamefont {C.-N.}\ \bibnamefont {Nie}}, \bibinfo {author} {\bibfnamefont {G.}~\bibnamefont {Zhou}}, \bibinfo {author} {\bibfnamefont {Z.}~\bibnamefont {Li}}, \bibinfo {author} {\bibfnamefont {W.}~\bibnamefont {Zhang}}, \bibinfo {author} {\bibfnamefont {C.-L.}\ \bibnamefont {Song}}, \bibinfo {author} {\bibfnamefont {K.}~\bibnamefont {He}}, \bibinfo {author} {\bibfnamefont {S.}~\bibnamefont {Ji}}, \bibinfo {author} {\bibfnamefont {S.}~\bibnamefont {Zhang}}, \bibinfo {author} {\bibfnamefont {L.}~\bibnamefont {Gu}}, \bibinfo {author} {\bibfnamefont {L.}~\bibnamefont {Wang}}, \bibinfo {author} {\bibfnamefont {X.-C.}\ \bibnamefont {Ma}}, \ and\ \bibinfo {author} {\bibfnamefont
  {Q.-K.}\ \bibnamefont {Xue}},\ }\bibfield  {title} {\enquote {\bibinfo {title} {Atomically resolved fese/srtio\textsubscript{3}(001) interface structure by scanning transmission electron microscopy},}\ }\href {\doibase 10.1088/2053-1583/3/2/024002} {\bibfield  {journal} {\bibinfo  {journal} {2D Materials}\ }\textbf {\bibinfo {volume} {3}},\ \bibinfo {pages} {024002} (\bibinfo {year} {2016})}\BibitemShut {NoStop}%
\bibitem [{\citenamefont {Peng}\ \emph {et~al.}(2020)\citenamefont {Peng}, \citenamefont {Zou}, \citenamefont {Han}, \citenamefont {Albright}, \citenamefont {Hong}, \citenamefont {Lau}, \citenamefont {Xu}, \citenamefont {Zhu}, \citenamefont {Walker},\ and\ \citenamefont {Ahn}}]{peng2020_picoscale}%
  \BibitemOpen
  \bibfield  {author} {\bibinfo {author} {\bibfnamefont {R.}~\bibnamefont {Peng}}, \bibinfo {author} {\bibfnamefont {K.}~\bibnamefont {Zou}}, \bibinfo {author} {\bibfnamefont {M.~G.}\ \bibnamefont {Han}}, \bibinfo {author} {\bibfnamefont {S.~D.}\ \bibnamefont {Albright}}, \bibinfo {author} {\bibfnamefont {H.}~\bibnamefont {Hong}}, \bibinfo {author} {\bibfnamefont {C.}~\bibnamefont {Lau}}, \bibinfo {author} {\bibfnamefont {H.~C.}\ \bibnamefont {Xu}}, \bibinfo {author} {\bibfnamefont {Y.}~\bibnamefont {Zhu}}, \bibinfo {author} {\bibfnamefont {F.~J.}\ \bibnamefont {Walker}}, \ and\ \bibinfo {author} {\bibfnamefont {C.~H.}\ \bibnamefont {Ahn}},\ }\bibfield  {title} {\enquote {\bibinfo {title} {Picoscale structural insight into superconductivity of monolayer {FeSe}/{SrTiO}$_{3}$},}\ }\href {\doibase 10.1126/sciadv.aay4517} {\bibfield  {journal} {\bibinfo  {journal} {Science Advances}\ }\textbf {\bibinfo {volume} {6}},\ \bibinfo {pages} {eaay4517} (\bibinfo {year} {2020})}\BibitemShut {NoStop}%
\bibitem [{\citenamefont {Yang}\ \emph {et~al.}(2024)\citenamefont {Yang}, \citenamefont {Zhang}, \citenamefont {Liu}, \citenamefont {Lu},\ and\ \citenamefont {Zhang}}]{yang2024_phonon}%
  \BibitemOpen
  \bibfield  {author} {\bibinfo {author} {\bibfnamefont {H.}~\bibnamefont {Yang}}, \bibinfo {author} {\bibfnamefont {P.}~\bibnamefont {Zhang}}, \bibinfo {author} {\bibfnamefont {K.}~\bibnamefont {Liu}}, \bibinfo {author} {\bibfnamefont {Z.}~\bibnamefont {Lu}}, \ and\ \bibinfo {author} {\bibfnamefont {S.}~\bibnamefont {Zhang}},\ }\bibfield  {title} {\enquote {\bibinfo {title} {Phonon modes and electron–phonon coupling at the fese/srtio$_3$ interface},}\ }\href {\doibase 10.1038/s41586-024-08118-0} {\bibfield  {journal} {\bibinfo  {journal} {Nature}\ }\textbf {\bibinfo {volume} {635}},\ \bibinfo {pages} {332--336} (\bibinfo {year} {2024})}\BibitemShut {NoStop}%
\bibitem [{\citenamefont {Wang}\ \emph {et~al.}(2016)\citenamefont {Wang}, \citenamefont {Linscheid}, \citenamefont {Berlijn},\ and\ \citenamefont {Johnston}}]{wang2016_epcoupling}%
  \BibitemOpen
  \bibfield  {author} {\bibinfo {author} {\bibfnamefont {Y.}~\bibnamefont {Wang}}, \bibinfo {author} {\bibfnamefont {A.}~\bibnamefont {Linscheid}}, \bibinfo {author} {\bibfnamefont {T.}~\bibnamefont {Berlijn}}, \ and\ \bibinfo {author} {\bibfnamefont {S.}~\bibnamefont {Johnston}},\ }\bibfield  {title} {\enquote {\bibinfo {title} {Ab initio study of cross-interface electron–phonon couplings in fese thin films on srtio$_3$ and batio$_3$},}\ }\href {\doibase 10.1103/PhysRevB.93.134513} {\bibfield  {journal} {\bibinfo  {journal} {Phys. Rev. B}\ }\textbf {\bibinfo {volume} {93}},\ \bibinfo {pages} {134513} (\bibinfo {year} {2016})}\BibitemShut {NoStop}%
\bibitem [{\citenamefont {Cao}\ \emph {et~al.}(2014)\citenamefont {Cao}, \citenamefont {Tan}, \citenamefont {Xiang},\ and\ \citenamefont {Gong}}]{cao2014_sdw}%
  \BibitemOpen
  \bibfield  {author} {\bibinfo {author} {\bibfnamefont {H.-Y.}\ \bibnamefont {Cao}}, \bibinfo {author} {\bibfnamefont {S.-S.}\ \bibnamefont {Tan}}, \bibinfo {author} {\bibfnamefont {H.}~\bibnamefont {Xiang}}, \ and\ \bibinfo {author} {\bibfnamefont {X.}~\bibnamefont {Gong}},\ }\bibfield  {title} {\enquote {\bibinfo {title} {Interfacial effects on the spin density wave in fese/srtio$_3$ thin films},}\ }\href {\doibase 10.1103/PhysRevB.89.014501} {\bibfield  {journal} {\bibinfo  {journal} {Phys. Rev. B}\ }\textbf {\bibinfo {volume} {89}},\ \bibinfo {pages} {014501} (\bibinfo {year} {2014})}\BibitemShut {NoStop}%
\bibitem [{\citenamefont {Gellé}\ \emph {et~al.}(2018)\citenamefont {Gellé}, \citenamefont {Chirita}, \citenamefont {Mertz}, \citenamefont {Rastei}, \citenamefont {Dinia},\ and\ \citenamefont {Colis}}]{gelle2018}%
  \BibitemOpen
  \bibfield  {author} {\bibinfo {author} {\bibfnamefont {F.}~\bibnamefont {Gellé}}, \bibinfo {author} {\bibfnamefont {R.}~\bibnamefont {Chirita}}, \bibinfo {author} {\bibfnamefont {D.}~\bibnamefont {Mertz}}, \bibinfo {author} {\bibfnamefont {M.-V.}\ \bibnamefont {Rastei}}, \bibinfo {author} {\bibfnamefont {A.}~\bibnamefont {Dinia}}, \ and\ \bibinfo {author} {\bibfnamefont {S.}~\bibnamefont {Colis}},\ }\bibfield  {title} {\enquote {\bibinfo {title} {Guideline to atomically flat tio\textsubscript{2}-terminated srtio\textsubscript{3}(001) surfaces},}\ }\href {\doibase 10.1016/j.susc.2018.06.001} {\bibfield  {journal} {\bibinfo  {journal} {Surface Science}\ }\textbf {\bibinfo {volume} {677}},\ \bibinfo {pages} {39--45} (\bibinfo {year} {2018})}\BibitemShut {NoStop}%
\bibitem [{Sup()}]{SuppMat}%
  \BibitemOpen
  \href@noop {} {}\bibinfo {note} {See Supplemental Material for AFM substrate characterization, asymmetric (112) X-ray diffraction scans of FeSe and FeTe, additional HAADF-STEM images of the lamella, and the one-dimensional Fourier-amplitude analysis used to extract the in-plane lattice constants from STEM line profiles.}\BibitemShut {Stop}%
\bibitem [{\citenamefont {Faeth}\ \emph {et~al.}(2021)\citenamefont {Faeth}, \citenamefont {Yang}, \citenamefont {Kawasaki}, \citenamefont {Nelson}, \citenamefont {Mishra}, \citenamefont {Chen}, \citenamefont {Schlom},\ and\ \citenamefont {Shen}}]{Faeth2021}%
  \BibitemOpen
  \bibfield  {author} {\bibinfo {author} {\bibfnamefont {B.~D.}\ \bibnamefont {Faeth}}, \bibinfo {author} {\bibfnamefont {S.}~\bibnamefont {Yang}}, \bibinfo {author} {\bibfnamefont {J.~K.}\ \bibnamefont {Kawasaki}}, \bibinfo {author} {\bibfnamefont {J.~N.}\ \bibnamefont {Nelson}}, \bibinfo {author} {\bibfnamefont {P.}~\bibnamefont {Mishra}}, \bibinfo {author} {\bibfnamefont {L.}~\bibnamefont {Chen}}, \bibinfo {author} {\bibfnamefont {D.~G.}\ \bibnamefont {Schlom}}, \ and\ \bibinfo {author} {\bibfnamefont {K.~M.}\ \bibnamefont {Shen}},\ }\bibfield  {title} {\enquote {\bibinfo {title} {Incoherent cooper pairing and pseudogap behavior in single-layer fese/srtio$_3$},}\ }\href {\doibase 10.1103/PhysRevX.11.021054} {\bibfield  {journal} {\bibinfo  {journal} {Physical Review X}\ }\textbf {\bibinfo {volume} {11}},\ \bibinfo {pages} {021054} (\bibinfo {year} {2021})}\BibitemShut {NoStop}%
\bibitem [{\citenamefont {Liu}\ and\ \citenamefont {Zou}(2020)}]{Liu2020}%
  \BibitemOpen
  \bibfield  {author} {\bibinfo {author} {\bibfnamefont {C.}~\bibnamefont {Liu}}\ and\ \bibinfo {author} {\bibfnamefont {K.}~\bibnamefont {Zou}},\ }\bibfield  {title} {\enquote {\bibinfo {title} {Tuning stoichiometry and its impact on superconductivity of monolayer and multilayer fese on srtio\textsubscript{3}},}\ }\href {\doibase 10.1103/PhysRevB.101.140502} {\bibfield  {journal} {\bibinfo  {journal} {Physical Review B}\ }\textbf {\bibinfo {volume} {101}},\ \bibinfo {pages} {140502(R)} (\bibinfo {year} {2020})}\BibitemShut {NoStop}%
\bibitem [{\citenamefont {Li}\ \emph {et~al.}(2021)\citenamefont {Li}, \citenamefont {Wang}, \citenamefont {Xiao}, \citenamefont {Li}, \citenamefont {Wang}, \citenamefont {Richardella}, \citenamefont {Wang},\ and\ \citenamefont {Samarth}}]{li2021_capping}%
  \BibitemOpen
  \bibfield  {author} {\bibinfo {author} {\bibfnamefont {Y.}~\bibnamefont {Li}}, \bibinfo {author} {\bibfnamefont {Z.}~\bibnamefont {Wang}}, \bibinfo {author} {\bibfnamefont {R.}~\bibnamefont {Xiao}}, \bibinfo {author} {\bibfnamefont {Q.}~\bibnamefont {Li}}, \bibinfo {author} {\bibfnamefont {K.}~\bibnamefont {Wang}}, \bibinfo {author} {\bibfnamefont {A.}~\bibnamefont {Richardella}}, \bibinfo {author} {\bibfnamefont {J.}~\bibnamefont {Wang}}, \ and\ \bibinfo {author} {\bibfnamefont {N.}~\bibnamefont {Samarth}},\ }\bibfield  {title} {\enquote {\bibinfo {title} {Capping layer influence and isotropic in-plane upper critical field of the superconductivity at the fese/srtio3 interface},}\ }\href {\doibase 10.1103/PhysRevMaterials.5.034802} {\bibfield  {journal} {\bibinfo  {journal} {Phys. Rev. Materials}\ }\textbf {\bibinfo {volume} {5}},\ \bibinfo {pages} {034802} (\bibinfo {year} {2021})}\BibitemShut {NoStop}%
\bibitem [{\citenamefont {Schneider}\ \emph {et~al.}(2014)\citenamefont {Schneider}, \citenamefont {Zaitsev}, \citenamefont {Fuchs},\ and\ \citenamefont {von Löhneysen}}]{Schneider2014}%
  \BibitemOpen
  \bibfield  {author} {\bibinfo {author} {\bibfnamefont {R.}~\bibnamefont {Schneider}}, \bibinfo {author} {\bibfnamefont {A.~G.}\ \bibnamefont {Zaitsev}}, \bibinfo {author} {\bibfnamefont {D.}~\bibnamefont {Fuchs}}, \ and\ \bibinfo {author} {\bibfnamefont {H.}~\bibnamefont {von Löhneysen}},\ }\bibfield  {title} {\enquote {\bibinfo {title} {Excess conductivity and {B}erezinskii–{K}osterlitz–{T}houless transition in superconducting {F}e{S}e thin films},}\ }\href {https://iopscience.iop.org/article/10.1088/0953-8984/26/45/455701} {\bibfield  {journal} {\bibinfo  {journal} {J. Phys. Condens. Matter}\ }\textbf {\bibinfo {volume} {26}},\ \bibinfo {pages} {455701} (\bibinfo {year} {2014})}\BibitemShut {NoStop}%
\bibitem [{\citenamefont {Zhao}\ \emph {et~al.}(2016)\citenamefont {Zhao}, \citenamefont {Chang}, \citenamefont {Xi}, \citenamefont {Mak},\ and\ \citenamefont {Moodera}}]{Zhao2016}%
  \BibitemOpen
  \bibfield  {author} {\bibinfo {author} {\bibfnamefont {W.}~\bibnamefont {Zhao}}, \bibinfo {author} {\bibfnamefont {C.-Z.}\ \bibnamefont {Chang}}, \bibinfo {author} {\bibfnamefont {X.}~\bibnamefont {Xi}}, \bibinfo {author} {\bibfnamefont {K.~F.}\ \bibnamefont {Mak}}, \ and\ \bibinfo {author} {\bibfnamefont {J.~S.}\ \bibnamefont {Moodera}},\ }\bibfield  {title} {\enquote {\bibinfo {title} {Vortex phase transitions in monolayer {F}e{S}e film on {S}r{T}i{O}3},}\ }\href {https://iopscience.iop.org/article/10.1088/2053-1583/3/2/024006} {\bibfield  {journal} {\bibinfo  {journal} {2D Mater.}\ }\textbf {\bibinfo {volume} {3}},\ \bibinfo {pages} {024006} (\bibinfo {year} {2016})}\BibitemShut {NoStop}%
\bibitem [{\citenamefont {Zhao}\ \emph {et~al.}(2024)\citenamefont {Zhao}, \citenamefont {Wei}, \citenamefont {Yu}, \citenamefont {Dong}, \citenamefont {Zhang}, \citenamefont {Chen}, \citenamefont {Li}, \citenamefont {Peng}, \citenamefont {Sun},\ and\ \citenamefont {Liu}}]{zhao2024_growth}%
  \BibitemOpen
  \bibfield  {author} {\bibinfo {author} {\bibfnamefont {Z.}~\bibnamefont {Zhao}}, \bibinfo {author} {\bibfnamefont {Z.}~\bibnamefont {Wei}}, \bibinfo {author} {\bibfnamefont {X.}~\bibnamefont {Yu}}, \bibinfo {author} {\bibfnamefont {W.}~\bibnamefont {Dong}}, \bibinfo {author} {\bibfnamefont {H.}~\bibnamefont {Zhang}}, \bibinfo {author} {\bibfnamefont {L.}~\bibnamefont {Chen}}, \bibinfo {author} {\bibfnamefont {J.}~\bibnamefont {Li}}, \bibinfo {author} {\bibfnamefont {Y.}~\bibnamefont {Peng}}, \bibinfo {author} {\bibfnamefont {Q.}~\bibnamefont {Sun}}, \ and\ \bibinfo {author} {\bibfnamefont {M.}~\bibnamefont {Liu}},\ }\bibfield  {title} {\enquote {\bibinfo {title} {Growth and characterization of high-quality fese thin films on srtio$_3$ with enhanced superconductivity},}\ }\href {\doibase 10.1103/PhysRevB.110.L140507} {\bibfield  {journal} {\bibinfo  {journal} {Phys. Rev. B}\ }\textbf {\bibinfo {volume} {110}},\ \bibinfo {pages} {L140507} (\bibinfo {year} {2024})}\BibitemShut {NoStop}%
\bibitem [{\citenamefont {Kosterlitz}\ and\ \citenamefont {Thouless}(1973)}]{Kosterlitz1973}%
  \BibitemOpen
  \bibfield  {author} {\bibinfo {author} {\bibfnamefont {J.~M.}\ \bibnamefont {Kosterlitz}}\ and\ \bibinfo {author} {\bibfnamefont {D.~J.}\ \bibnamefont {Thouless}},\ }\bibfield  {title} {\enquote {\bibinfo {title} {Ordering, metastability and phase transitions in two-dimensional systems},}\ }\href {https://iopscience.iop.org/article/10.1088/0022-3719/6/7/010} {\bibfield  {journal} {\bibinfo  {journal} {J. Phys. C Sol. St. Phys.}\ }\textbf {\bibinfo {volume} {6}},\ \bibinfo {pages} {1181} (\bibinfo {year} {1973})}\BibitemShut {NoStop}%
\bibitem [{\citenamefont {Halperin}\ and\ \citenamefont {Nelson}(1979)}]{Halperin1979}%
  \BibitemOpen
  \bibfield  {author} {\bibinfo {author} {\bibfnamefont {B.~I.}\ \bibnamefont {Halperin}}\ and\ \bibinfo {author} {\bibfnamefont {D.~R.}\ \bibnamefont {Nelson}},\ }\bibfield  {title} {\enquote {\bibinfo {title} {Resistive transition in superconducting films},}\ }\href {https://link.springer.com/article/10.1007/BF00116988} {\bibfield  {journal} {\bibinfo  {journal} {J. Low Temp. Phys.}\ }\textbf {\bibinfo {volume} {36}},\ \bibinfo {pages} {599--616} (\bibinfo {year} {1979})}\BibitemShut {NoStop}%
\bibitem [{\citenamefont {Miller}\ \emph {et~al.}(2022)\citenamefont {Miller}, \citenamefont {Wiebe},\ and\ \citenamefont {Dutton}}]{miller2022_laue_oscillations}%
  \BibitemOpen
  \bibfield  {author} {\bibinfo {author} {\bibfnamefont {A.~M.}\ \bibnamefont {Miller}}, \bibinfo {author} {\bibfnamefont {C.~R.}\ \bibnamefont {Wiebe}}, \ and\ \bibinfo {author} {\bibfnamefont {S.~E.}\ \bibnamefont {Dutton}},\ }\bibfield  {title} {\enquote {\bibinfo {title} {Extracting information from x-ray diffraction patterns containing laue oscillations},}\ }\href {\doibase 10.1515/znb-2022-0020} {\bibfield  {journal} {\bibinfo  {journal} {Zeitschrift für Naturforschung B}\ }\textbf {\bibinfo {volume} {77}},\ \bibinfo {pages} {385--392} (\bibinfo {year} {2022})}\BibitemShut {NoStop}%
\bibitem [{\citenamefont {Huang}\ \emph {et~al.}(2021)\citenamefont {Huang}, \citenamefont {Chen},\ and\ \citenamefont {Lei}}]{huang2021_directARPES}%
  \BibitemOpen
  \bibfield  {author} {\bibinfo {author} {\bibfnamefont {G.-Q.}\ \bibnamefont {Huang}}, \bibinfo {author} {\bibfnamefont {B.}~\bibnamefont {Chen}}, \ and\ \bibinfo {author} {\bibfnamefont {H.}~\bibnamefont {Lei}},\ }\bibfield  {title} {\enquote {\bibinfo {title} {Direct-arpes and stm investigation of fese thin film growth by nd:yag laser},}\ }\href {\doibase 10.3390/coatings11030276} {\bibfield  {journal} {\bibinfo  {journal} {Coatings}\ }\textbf {\bibinfo {volume} {11}},\ \bibinfo {pages} {276} (\bibinfo {year} {2021})}\BibitemShut {NoStop}%
\bibitem [{\citenamefont {Feng}\ \emph {et~al.}(2018)\citenamefont {Feng}, \citenamefont {Yuan}, \citenamefont {He}, \citenamefont {Hu}, \citenamefont {Lin}, \citenamefont {Li}, \citenamefont {Jiang}, \citenamefont {Huang}, \citenamefont {Ni}, \citenamefont {Li}, \citenamefont {Zhu}, \citenamefont {Dong}, \citenamefont {Zhou}, \citenamefont {Wang}, \citenamefont {Zhao},\ and\ \citenamefont {Jin}}]{feng2018_tunable}%
  \BibitemOpen
  \bibfield  {author} {\bibinfo {author} {\bibfnamefont {Z.}~\bibnamefont {Feng}}, \bibinfo {author} {\bibfnamefont {J.}~\bibnamefont {Yuan}}, \bibinfo {author} {\bibfnamefont {G.}~\bibnamefont {He}}, \bibinfo {author} {\bibfnamefont {W.}~\bibnamefont {Hu}}, \bibinfo {author} {\bibfnamefont {Z.}~\bibnamefont {Lin}}, \bibinfo {author} {\bibfnamefont {D.}~\bibnamefont {Li}}, \bibinfo {author} {\bibfnamefont {X.}~\bibnamefont {Jiang}}, \bibinfo {author} {\bibfnamefont {Y.}~\bibnamefont {Huang}}, \bibinfo {author} {\bibfnamefont {S.}~\bibnamefont {Ni}}, \bibinfo {author} {\bibfnamefont {J.}~\bibnamefont {Li}}, \bibinfo {author} {\bibfnamefont {B.}~\bibnamefont {Zhu}}, \bibinfo {author} {\bibfnamefont {X.}~\bibnamefont {Dong}}, \bibinfo {author} {\bibfnamefont {F.}~\bibnamefont {Zhou}}, \bibinfo {author} {\bibfnamefont {H.}~\bibnamefont {Wang}}, \bibinfo {author} {\bibfnamefont {Z.}~\bibnamefont {Zhao}}, \ and\ \bibinfo {author} {\bibfnamefont {K.}~\bibnamefont {Jin}},\ }\bibfield  {title} {\enquote {\bibinfo
  {title} {Tunable critical temperature for superconductivity in fese thin films by pulsed laser deposition},}\ }\href {\doibase 10.1038/s41598-018-22291-z} {\bibfield  {journal} {\bibinfo  {journal} {Scientific Reports}\ }\textbf {\bibinfo {volume} {8}},\ \bibinfo {pages} {4039} (\bibinfo {year} {2018})}\BibitemShut {NoStop}%
\bibitem [{\citenamefont {Liu}\ and\ \citenamefont {Wang}(2020)}]{liu2020_fe_se_2dmaterials}%
  \BibitemOpen
  \bibfield  {author} {\bibinfo {author} {\bibfnamefont {C.}~\bibnamefont {Liu}}\ and\ \bibinfo {author} {\bibfnamefont {J.}~\bibnamefont {Wang}},\ }\bibfield  {title} {\enquote {\bibinfo {title} {Heterostructural one-unit-cell fese/srtio$_3$: from high-temperature superconductivity to topological states},}\ }\href {\doibase 10.1088/2053-1583/ab734b} {\bibfield  {journal} {\bibinfo  {journal} {2D Materials}\ }\textbf {\bibinfo {volume} {7}},\ \bibinfo {pages} {022006} (\bibinfo {year} {2020})}\BibitemShut {NoStop}%
\bibitem [{\citenamefont {Peng}\ \emph {et~al.}(2014{\natexlab{b}})\citenamefont {Peng}, \citenamefont {Shen}, \citenamefont {Xie}, \citenamefont {Xu}, \citenamefont {Tan}, \citenamefont {Xia}, \citenamefont {Zhang}, \citenamefont {Cao}, \citenamefont {Gong}, \citenamefont {Hu}, \citenamefont {Xie},\ and\ \citenamefont {Feng}}]{peng2014_tuning}%
  \BibitemOpen
  \bibfield  {author} {\bibinfo {author} {\bibfnamefont {R.}~\bibnamefont {Peng}}, \bibinfo {author} {\bibfnamefont {X.~P.}\ \bibnamefont {Shen}}, \bibinfo {author} {\bibfnamefont {X.~X.}\ \bibnamefont {Xie}}, \bibinfo {author} {\bibfnamefont {H.~C.}\ \bibnamefont {Xu}}, \bibinfo {author} {\bibfnamefont {S.~Y.}\ \bibnamefont {Tan}}, \bibinfo {author} {\bibfnamefont {M.~Y.}\ \bibnamefont {Xia}}, \bibinfo {author} {\bibfnamefont {T.}~\bibnamefont {Zhang}}, \bibinfo {author} {\bibfnamefont {H.~Y.}\ \bibnamefont {Cao}}, \bibinfo {author} {\bibfnamefont {X.~G.}\ \bibnamefont {Gong}}, \bibinfo {author} {\bibfnamefont {J.~P.}\ \bibnamefont {Hu}}, \bibinfo {author} {\bibfnamefont {B.~P.}\ \bibnamefont {Xie}}, \ and\ \bibinfo {author} {\bibfnamefont {D.~L.}\ \bibnamefont {Feng}},\ }\bibfield  {title} {\enquote {\bibinfo {title} {Tuning the band structure and superconductivity in single-layer fese by interface engineering},}\ }\href {\doibase 10.1038/ncomms6044} {\bibfield  {journal} {\bibinfo  {journal} {Nature
  Communications}\ }\textbf {\bibinfo {volume} {5}},\ \bibinfo {pages} {5044} (\bibinfo {year} {2014}{\natexlab{b}})}\BibitemShut {NoStop}%
\bibitem [{\citenamefont {Tan}\ \emph {et~al.}(2013)\citenamefont {Tan}, \citenamefont {Xia}, \citenamefont {Zhang}, \citenamefont {Ye}, \citenamefont {Chen}, \citenamefont {Xie}, \citenamefont {Peng}, \citenamefont {Xu}, \citenamefont {Fan}, \citenamefont {Xu} \emph {et~al.}}]{tan2013_interface}%
  \BibitemOpen
  \bibfield  {author} {\bibinfo {author} {\bibfnamefont {S.~Y.}\ \bibnamefont {Tan}}, \bibinfo {author} {\bibfnamefont {M.~Y.}\ \bibnamefont {Xia}}, \bibinfo {author} {\bibfnamefont {Y.}~\bibnamefont {Zhang}}, \bibinfo {author} {\bibfnamefont {Z.~R.}\ \bibnamefont {Ye}}, \bibinfo {author} {\bibfnamefont {F.}~\bibnamefont {Chen}}, \bibinfo {author} {\bibfnamefont {X.~X.}\ \bibnamefont {Xie}}, \bibinfo {author} {\bibfnamefont {R.}~\bibnamefont {Peng}}, \bibinfo {author} {\bibfnamefont {D.~F.}\ \bibnamefont {Xu}}, \bibinfo {author} {\bibfnamefont {Q.}~\bibnamefont {Fan}}, \bibinfo {author} {\bibfnamefont {H.~C.}\ \bibnamefont {Xu}},  \emph {et~al.},\ }\bibfield  {title} {\enquote {\bibinfo {title} {Interface‑induced superconductivity and strain‑dependent spin density wave in fese/srtio$_3$ thin films},}\ }\href {\doibase 10.1038/nmat3654} {\bibfield  {journal} {\bibinfo  {journal} {Nature Materials}\ }\textbf {\bibinfo {volume} {12}},\ \bibinfo {pages} {634--640} (\bibinfo {year} {2013})}\BibitemShut
  {NoStop}%
\bibitem [{\citenamefont {Erdman}\ \emph {et~al.}(2002)\citenamefont {Erdman}, \citenamefont {Poeppelmeier}, \citenamefont {Asta}, \citenamefont {Warschkow}, \citenamefont {Ellis},\ and\ \citenamefont {Marks}}]{erdman2002_tio2}%
  \BibitemOpen
  \bibfield  {author} {\bibinfo {author} {\bibfnamefont {N.}~\bibnamefont {Erdman}}, \bibinfo {author} {\bibfnamefont {K.}~\bibnamefont {Poeppelmeier}}, \bibinfo {author} {\bibfnamefont {M.}~\bibnamefont {Asta}}, \bibinfo {author} {\bibfnamefont {O.}~\bibnamefont {Warschkow}}, \bibinfo {author} {\bibfnamefont {D.}~\bibnamefont {Ellis}}, \ and\ \bibinfo {author} {\bibfnamefont {L.}~\bibnamefont {Marks}},\ }\bibfield  {title} {\enquote {\bibinfo {title} {The structure and chemistry of the tio\textsubscript{2}-rich surface of srtio\textsubscript{3}(001)},}\ }\href {\doibase https://doi.org/10.1038/nature01010} {\bibfield  {journal} {\bibinfo  {journal} {Nature}\ }\textbf {\bibinfo {volume} {419}},\ \bibinfo {pages} {55--58} (\bibinfo {year} {2002})}\BibitemShut {NoStop}%
\bibitem [{\citenamefont {Herger}\ \emph {et~al.}(2007)\citenamefont {Herger}, \citenamefont {Willmott}, \citenamefont {Bunk}, \citenamefont {Schlepuetz}, \citenamefont {Patterson},\ and\ \citenamefont {Delley}}]{herger2007_srtio3}%
  \BibitemOpen
  \bibfield  {author} {\bibinfo {author} {\bibfnamefont {R.}~\bibnamefont {Herger}}, \bibinfo {author} {\bibfnamefont {P.~R.}\ \bibnamefont {Willmott}}, \bibinfo {author} {\bibfnamefont {O.}~\bibnamefont {Bunk}}, \bibinfo {author} {\bibfnamefont {C.~M.}\ \bibnamefont {Schlepuetz}}, \bibinfo {author} {\bibfnamefont {B.~D.}\ \bibnamefont {Patterson}}, \ and\ \bibinfo {author} {\bibfnamefont {B.}~\bibnamefont {Delley}},\ }\bibfield  {title} {\enquote {\bibinfo {title} {Surface of strontium titanate (001)},}\ }\href {\doibase 10.1103/PhysRevLett.98.076102} {\bibfield  {journal} {\bibinfo  {journal} {Phys. Rev. Lett.}\ }\textbf {\bibinfo {volume} {98}},\ \bibinfo {pages} {076102} (\bibinfo {year} {2007})}\BibitemShut {NoStop}%
\bibitem [{\citenamefont {Zhu}\ \emph {et~al.}(2012)\citenamefont {Zhu}, \citenamefont {Radtke},\ and\ \citenamefont {Botton}}]{zhu2012_reconstruction}%
  \BibitemOpen
  \bibfield  {author} {\bibinfo {author} {\bibfnamefont {G.-Z.}\ \bibnamefont {Zhu}}, \bibinfo {author} {\bibfnamefont {G.}~\bibnamefont {Radtke}}, \ and\ \bibinfo {author} {\bibfnamefont {G.~A.}\ \bibnamefont {Botton}},\ }\bibfield  {title} {\enquote {\bibinfo {title} {Bonding and structure of a reconstructed (001) surface of srtio$_3$ from tem},}\ }\href {\doibase 10.1038/nature11563} {\bibfield  {journal} {\bibinfo  {journal} {Nature}\ }\textbf {\bibinfo {volume} {490}},\ \bibinfo {pages} {384--387} (\bibinfo {year} {2012})}\BibitemShut {NoStop}%
\bibitem [{\citenamefont {Kienzle}\ \emph {et~al.}(2011)\citenamefont {Kienzle}, \citenamefont {Becerra-Toledo},\ and\ \citenamefont {Marks}}]{kienzle2011_vacant}%
  \BibitemOpen
  \bibfield  {author} {\bibinfo {author} {\bibfnamefont {D.~M.}\ \bibnamefont {Kienzle}}, \bibinfo {author} {\bibfnamefont {A.~E.}\ \bibnamefont {Becerra-Toledo}}, \ and\ \bibinfo {author} {\bibfnamefont {L.~D.}\ \bibnamefont {Marks}},\ }\bibfield  {title} {\enquote {\bibinfo {title} {Vacant-site octahedral tilings on srtio$_3$ (001), the $(\sqrt{13} \times \sqrt{13})r33.7^\circ$ surface, and related structures},}\ }\href {\doibase 10.1103/PhysRevLett.106.176102} {\bibfield  {journal} {\bibinfo  {journal} {Phys. Rev. Lett.}\ }\textbf {\bibinfo {volume} {106}},\ \bibinfo {pages} {176102} (\bibinfo {year} {2011})}\BibitemShut {NoStop}%
\bibitem [{\citenamefont {Marshall}\ \emph {et~al.}(2015)\citenamefont {Marshall}, \citenamefont {Becerra-Toledo}, \citenamefont {Marks},\ and\ \citenamefont {Castell}}]{marshall2015_defects}%
  \BibitemOpen
  \bibfield  {author} {\bibinfo {author} {\bibfnamefont {M.~S.~J.}\ \bibnamefont {Marshall}}, \bibinfo {author} {\bibfnamefont {A.~E.}\ \bibnamefont {Becerra-Toledo}}, \bibinfo {author} {\bibfnamefont {L.~D.}\ \bibnamefont {Marks}}, \ and\ \bibinfo {author} {\bibfnamefont {M.~R.}\ \bibnamefont {Castell}},\ }\bibfield  {title} {\enquote {\bibinfo {title} {Defects on strontium titanate},}\ }in\ \href {\doibase 10.1007/978-3-319-14367-5_11} {\emph {\bibinfo {booktitle} {Defects at Oxide Surfaces}}},\ \bibinfo {series} {Springer Series in Surface Sciences}, Vol.~\bibinfo {volume} {58},\ \bibinfo {editor} {edited by\ \bibinfo {editor} {\bibfnamefont {J.}~\bibnamefont {Jupille}}\ and\ \bibinfo {editor} {\bibfnamefont {G.}~\bibnamefont {Thornton}}}\ (\bibinfo  {publisher} {Springer International Publishing Switzerland},\ \bibinfo {year} {2015})\ pp.\ \bibinfo {pages} {327--349}\BibitemShut {NoStop}%
\bibitem [{\citenamefont {Jeschke}\ \emph {et~al.}(2015)\citenamefont {Jeschke}, \citenamefont {Shen},\ and\ \citenamefont {Valenti}}]{jeschke2015oxygen}%
  \BibitemOpen
  \bibfield  {author} {\bibinfo {author} {\bibfnamefont {Harald~O.}\ \bibnamefont {Jeschke}}, \bibinfo {author} {\bibfnamefont {Juan}\ \bibnamefont {Shen}}, \ and\ \bibinfo {author} {\bibfnamefont {Roser}\ \bibnamefont {Valenti}},\ }\bibfield  {title} {\enquote {\bibinfo {title} {Oxygen vacancies in srtio$_3$ and their impact on electronic structure},}\ }\href {\doibase 10.1088/1367-2630/17/2/023034} {\bibfield  {journal} {\bibinfo  {journal} {New Journal of Physics}\ }\textbf {\bibinfo {volume} {17}},\ \bibinfo {pages} {023034} (\bibinfo {year} {2015})}\BibitemShut {NoStop}%
\bibitem [{\citenamefont {Koma}(1992)}]{Koma1992}%
  \BibitemOpen
  \bibfield  {author} {\bibinfo {author} {\bibfnamefont {A.}~\bibnamefont {Koma}},\ }\bibfield  {title} {\enquote {\bibinfo {title} {Van der {W}aals epitaxy—a new epitaxial growth method for a highly lattice-mismatched system},}\ }\href {https://www.sciencedirect.com/science/article/pii/0040609092908729} {\bibfield  {journal} {\bibinfo  {journal} {Thin Solid Films}\ }\textbf {\bibinfo {volume} {216}},\ \bibinfo {pages} {72--76} (\bibinfo {year} {1992})}\BibitemShut {NoStop}%
\bibitem [{\citenamefont {Walsh}\ and\ \citenamefont {Hinkle}(2017)}]{Walsh2017}%
  \BibitemOpen
  \bibfield  {author} {\bibinfo {author} {\bibfnamefont {L.~A.}\ \bibnamefont {Walsh}}\ and\ \bibinfo {author} {\bibfnamefont {C.~L.}\ \bibnamefont {Hinkle}},\ }\bibfield  {title} {\enquote {\bibinfo {title} {Van der {W}aals epitaxy: 2{D} materials and topological insulators},}\ }\href {https://www.sciencedirect.com/science/article/pii/S2352940717302822} {\bibfield  {journal} {\bibinfo  {journal} {Appl. Mater. Today}\ }\textbf {\bibinfo {volume} {9}},\ \bibinfo {pages} {504--515} (\bibinfo {year} {2017})}\BibitemShut {NoStop}%
\end{thebibliography}


\onecolumngrid

%\hrulefill

\clearpage

\large
\centerline{\textbf{Unconventional incommensurate epitaxy of superconducting FeSe films on SrTiO\textsubscript{3}}}
\centerline{\textbf{- Supplemental Materials}}
\vspace{1em}
\normalsize
\centerline{M. Klement,$^{1,2}$ K. M. Fijalkowski,$^{1,2}$ M. Kamp,$^{3}$ C. Gould,$^{1,2}$ and L. W. Molenkamp$^{1,2}$}
\vspace{1em}
\bigskip

\twocolumngrid

\setcounter{figure}{0} 
\renewcommand{\figurename}{Fig. S\!\!}

\section[1. MBE growth]{1. MBE growth}

\subsection{Fig. S1}
The AFM image in Fig. S1 shows the surface morphology of a SrTiO\textsubscript{3}(001) substrate after buffered HF etching and high-temperature O\textsubscript{2} annealing, yielding atomically flat terraces with single-unit-cell step heights ($\sim$0.4 nm). Such a well-ordered TiO\textsubscript{2} termination is crucial for obtaining the registry-free yet epitaxially aligned FeSe growth reported in this work, as it minimizes uncontrolled nucleation sites and ensures a uniform in-plane orientation. The terrace quality shown here is representative of the substrates used throughout our experiments.

\begin{figure}[ht]
  \centering
  \includegraphics[width=1\textwidth]{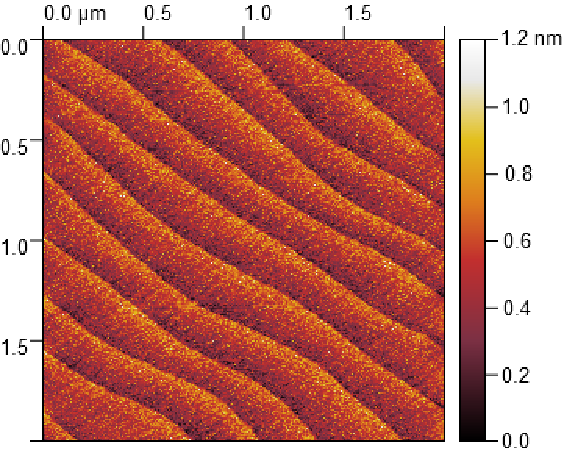}
  \caption{Atomic force microscopy (AFM) topography of a SrTiO\textsubscript{3}(001) substrate after chemical etching and high-temperature annealing. The image reveals an array of well-defined step-terrace structures with a uniform step height of approximately one unit cell ($\sim$0.4 nm), consistent with the expected TiO\textsubscript{2} termination. The terrace morphology confirms the effectiveness of the surface preparation procedure and the formation of atomically flat TiO\textsubscript{2}-terminated surfaces suitable for epitaxial FeSe growth.}
  \label{fig:S1}
\end{figure}

\section[2. X-ray Diffraction Analysis]{2. X-ray Diffraction Analysis}
\subsection{Fig. S2}
The XRD scan of the FeSe(002) reflection in Fig. S2 confirms the epitaxial growth of the FeSe layer on SrTiO\textsubscript{3}(001) with the c-axis oriented perpendicular to the substrate. The presence of a single, weak Kiessig fringe on the low-angle side indicates a well-defined but comparatively thin layer, with a thickness of about 4.2 nm as derived from fringe spacing. In comparison to the FeTe layer, where multiple fringes with high visibility are observed, the reduced fringe contrast here is attributed primarily to the lower scattering volume of the thinner FeSe film and an increased relative noise contribution. The peak position further confirms that the FeSe layer retains its intrinsic c-axis lattice parameter rather than adopting that of the substrate, in agreement with the registry-free epitaxy described in the main text.

\begin{figure}[ht]
  \centering
  \includegraphics[width=1\textwidth]{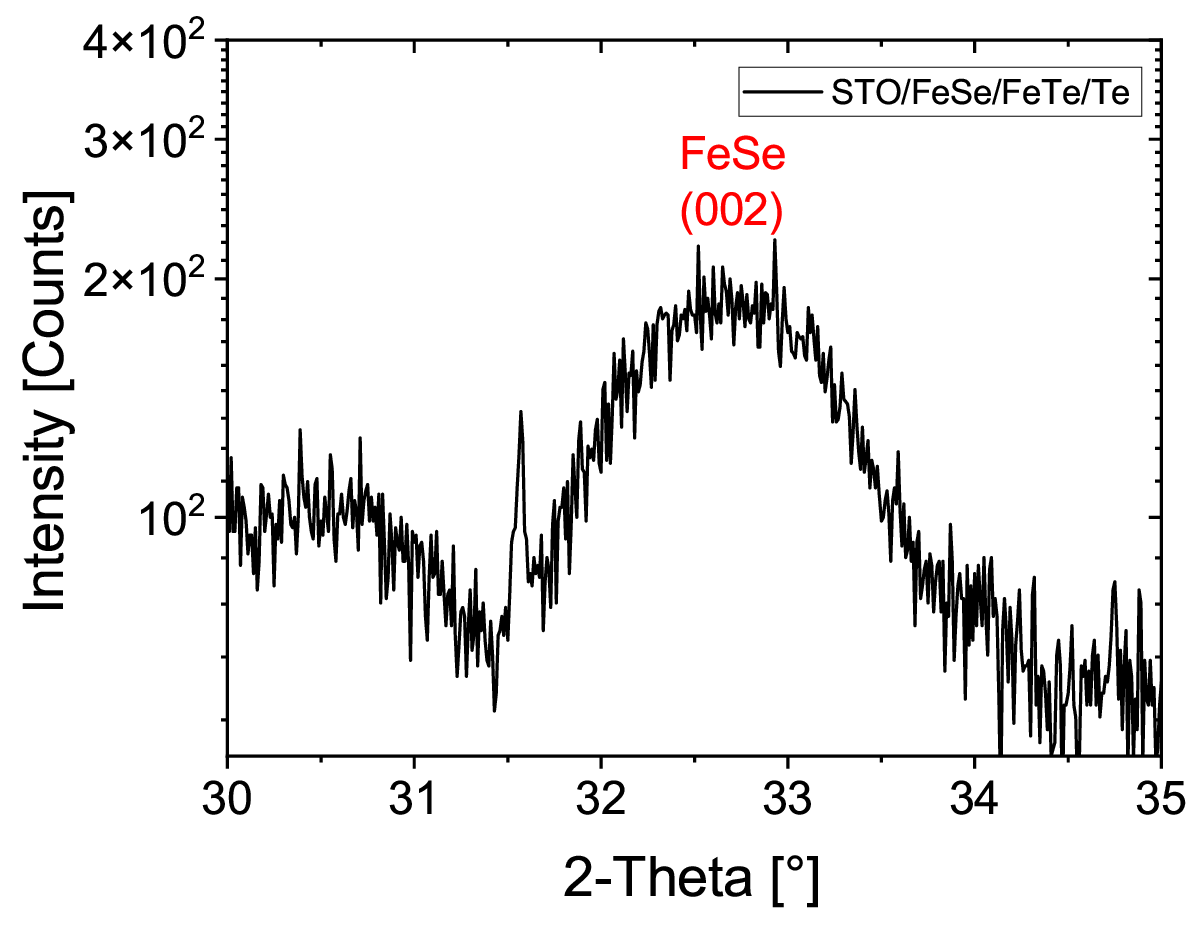}
  \caption{X-ray diffraction scan around the FeSe(002) reflection. A single weak Kiessig fringe is visible on the low-angle side (left of the Bragg peak), suggesting a film thickness of approximately 4.2 nm. In contrast to the FeTe layer, where multiple fringes can be observed (cf. Figure 3a), only one fringe is discernible here, resulting in increased uncertainty. This difference is attributed to the thinner FeSe layer, reduced signal strength, and elevated noise level.}
  \label{fig:S2}
\end{figure}

\subsection{Fig. S3}
The asymmetric XRD scan of the FeSe(112) reflection, measured in tilted geometry, provides direct access to the in-plane lattice constant. The Voigt fit to the diffraction peak yields an a-parameter of 3.862 \AA, which deviates from the bulk FeSe value but remains clearly distinct from the SrTiO\textsubscript{3} in-plane lattice constant. Such deviations from bulk values are commonly observed in nanometer-thick films and do not imply pseudomorphic growth. Instead, the mismatch with SrTiO\textsubscript{3} demonstrates that FeSe retains its intrinsic lattice dimensions rather than adopting those of the substrate. The consistency of this result across multiple samples further confirms the robustness of the registry-free epitaxy discussed in the main text.

\begin{figure}[ht]
  \centering
  \includegraphics[width=1\textwidth]{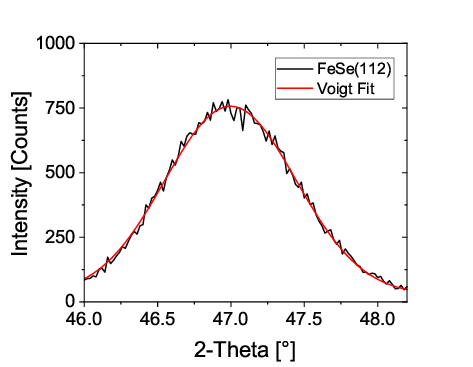}
  \caption{Asymmetric X-ray diffraction scan of the FeSe(112) reflection. The measurement is performed in a tilted geometry to access the in-plane lattice information. From the peak position and assuming tetragonal symmetry, an in-plane lattice parameter of (a = 3.862 \AA) is extracted.}
  \label{fig:S3}
\end{figure}

\subsection{Fig. S4}
The asymmetric XRD scan of the FeTe(112) reflection reveals an in-plane lattice constant of 3.824 \AA, as determined from the Bragg equation under the assumption of tetragonal symmetry. While this value differs slightly from that of FeSe ($\sim$0.02 \AA), the STEM images show atomically sharp interfaces without any detectable strain contrast or interfacial defects. This indicates that FeTe and FeSe form a structurally continuous, orientation-aligned interface despite their small in-plane lattice mismatch, consistent with the registry-free, incommensurate epitaxial relationship discussed in the main text.

\begin{figure}[ht]
  \centering
  \includegraphics[width=1\textwidth]{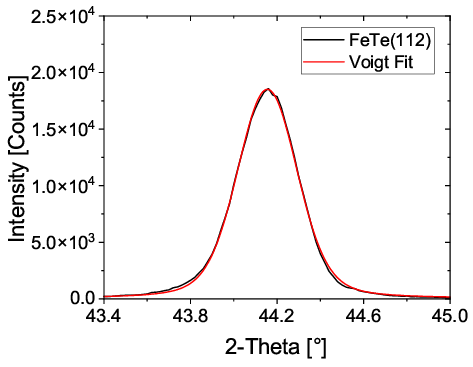}
  \caption{Asymmetric X-ray diffraction scan of the FeTe(112) reflection. The tilted measurement allows determination of the in-plane lattice constant. Using the Bragg equation and assuming tetragonal symmetry, the extracted in-plane lattice parameter is (a = 3.824 \AA).}
  \label{fig:S4}
\end{figure}

\subsection[3. Transmission Electron Microscopy
Analysis]{3. Transmission Electron Microscopy
Analysis}
\subsection{Fig. S5}
The STEM image in Fig. S5 shows the cross-sectional lamella prepared by focused ion beam (FIB) milling from a representative surface region. The lamella has a total length of $\sim$8 $\mu$m and was thinned to electron transparency for high-resolution STEM. 

\begin{figure}[ht]
  \centering
  \includegraphics[width=\linewidth]{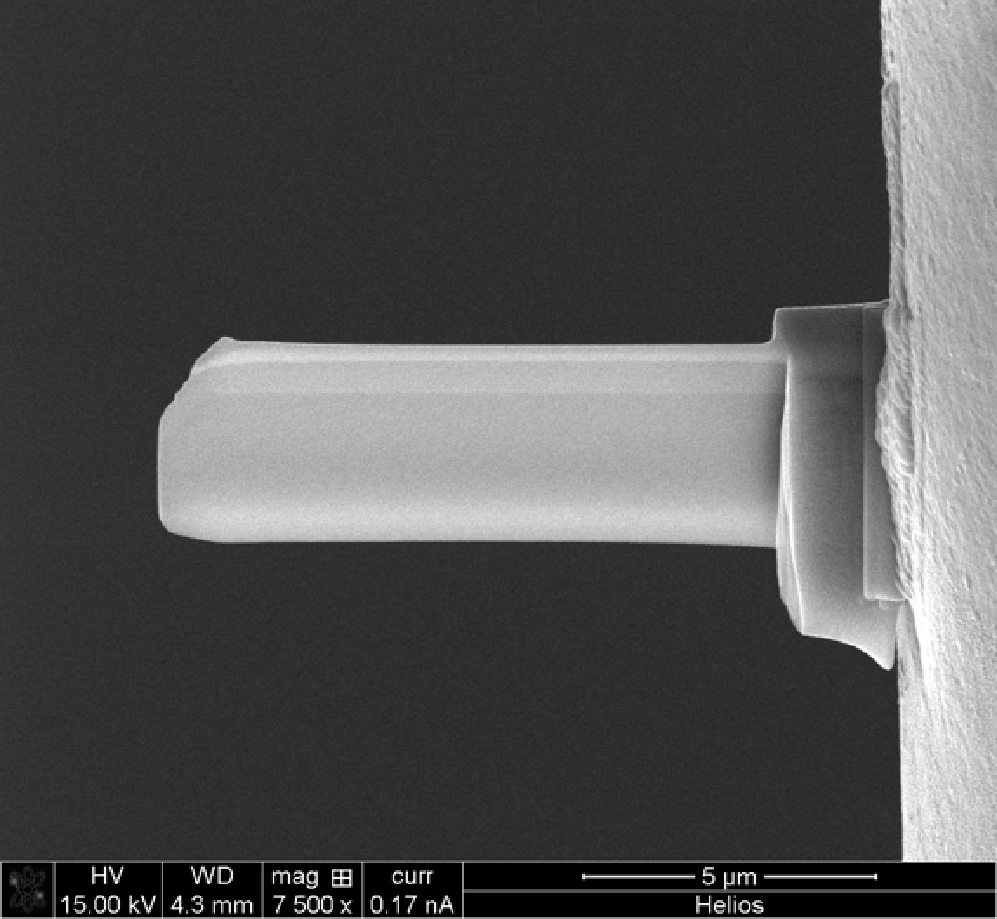}
  \caption{The STEM image shows a cross-sectional lamella ($\sim$8 $\mu$m) prepared by FIB and thinned to electron transparency. Imaging along $\sim$25\% of its length confirmed laterally uniform, registry-free epitaxy with atomically sharp interfaces and no misfit dislocations or extended defects, consistent with the XRD results.}
  \label{fig:S5}
\end{figure}

Multiple regions distributed along the lamella were imaged at various magnifications—together covering roughly 25\% of its total length. This sampling strategy ensured that our structural assessment was not confined to a single local area and allowed us to confirm (i) lateral uniformity of the registry-free epitaxy, (ii) atomically abrupt FeSe/SrTiO\textsubscript{3} and FeTe/FeSe interfaces, and (iii) the absence of misfit dislocations, stacking faults, or amorphous inclusions within the surveyed area. These real-space observations complement the XRD analysis presented in the main text.

\subsection{Fig. S6 \& S7}
Figures S6 and S7 present complementary STEM images of the SrTiO\textsubscript{3}(001)/FeSe/FeTe heterostructure, acquired from different regions of the same FIB-prepared lamella. The medium-magnification image in Fig. S6 highlights the overall structural integrity of the multilayer stack, with uniform layering and abrupt interfaces across several nanometers. 

In contrast, the high-resolution view in Fig. S7 resolves individual atomic columns, confirming the crystalline order and sharpness of the interfaces at the atomic scale. Together, these images demonstrate the absence of misfit dislocations or amorphous interlayers and are fully consistent with the registry-free epitaxy inferred from XRD.

\begin{figure}[ht]
  \centering
  \includegraphics[width=\linewidth]{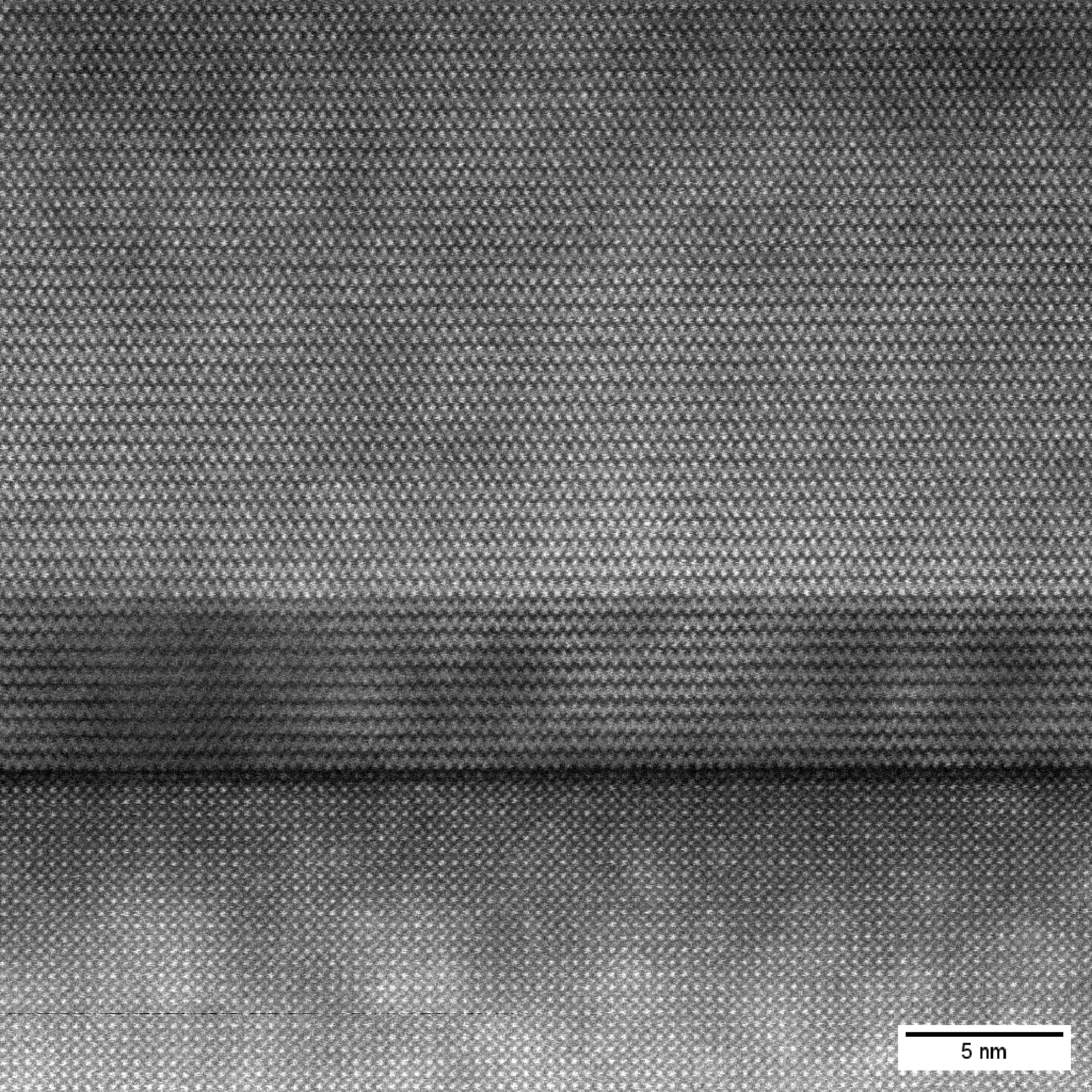}
  \caption{HAADF-STEM image of the FeSe/FeTe multilayer on SrTiO\textsubscript{3}(001) recorded at medium magnification (scale bar: 5 nm). The FIB-milled lamella was thinned to electron transparency. The image illustrates uniform layering and atomically abrupt interfaces in the field of view; no misfit dislocations are detected.}
  \label{fig:S6}
\end{figure}

\begin{figure}[ht]
  \centering
  \includegraphics[width=\linewidth]{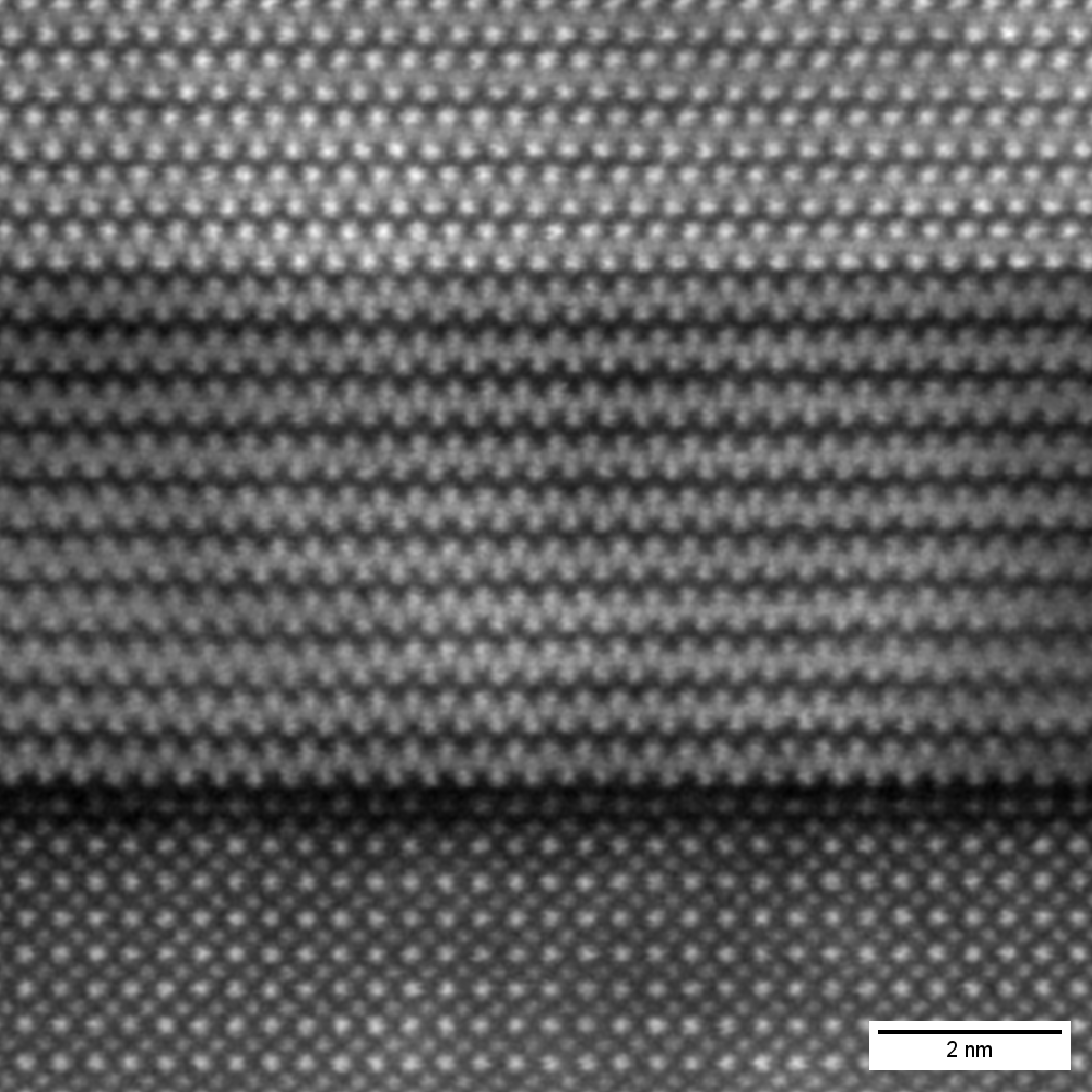}
  \caption{HAADF-STEM image acquired at a different location along the same lamella as in Fig. S6 (scale bar: 2 nm). Atomic columns are well resolved, and the interfaces within the field of view remain atomically sharp and defect-free, without any indication of interfacial strain contrast or lateral disorder.
}
  \label{fig:S7}
\end{figure}

\subsection{Fig. S8}
Atomic-row intensity traces were extracted directly at each of the three relevant 
heterointerfaces: on the uppermost SrO layer of SrTiO\textsubscript{3}, on the adjacent first FeSe layer at the STO--FeSe interface, and on the lowest FeTe layer at the FeTe--FeSe interface. Panel (a) illustrates this procedure by showing one sample trace at the STO--FeSe interface; analogous traces were acquired in the same manner for FeSe and FeTe at their respective interfaces.

The resulting real-space intensity profiles were evaluated to track the lateral atomic 
periodicity across the interface region. At the STO--FeSe interface, the two corresponding profiles (panel b) exhibit an initially aligned periodicity that gradually develops a systematic phase shift over roughly 60 unit cells, consistent with an in-plane mismatch accumulation.

Each profile $I(x)$ was then numerically converted into a one-dimensional Fourier 
amplitude spectrum $|F(k)|$ (panel c), yielding the first-harmonic peaks of STO, FeSe, and FeTe. These peaks were fitted with Gaussian functions to determine their precise spatial frequencies. The SrTiO\textsubscript{3} peak was fixed to the known in-plane lattice spacing of $a_{\mathrm{STO}} = 3.905$ \AA, providing the calibration factor required to convert 
pixel-based frequencies into physical units, from which the in-plane lattice constants of FeSe and FeTe were obtained.

Applying this procedure to eleven independent STEM images of various lamella regions produced weighted averages of $a_{\mathrm{FeSe}} = 3.843 \pm 0.012$ \AA and 
$a_{\mathrm{FeTe}} = 3.826 \pm 0.010$~\AA. These values agree with the XRD-extracted lattice constants and confirm that neither FeSe nor FeTe adopts the SrTiO\textsubscript{3} in-plane lattice parameter, supporting the interpretation of a registry-free, incommensurate epitaxial relationship evident in the Fourier amplitude signatures (panel c).

\begin{figure}[ht]
  \centering
  \includegraphics[width=\linewidth]{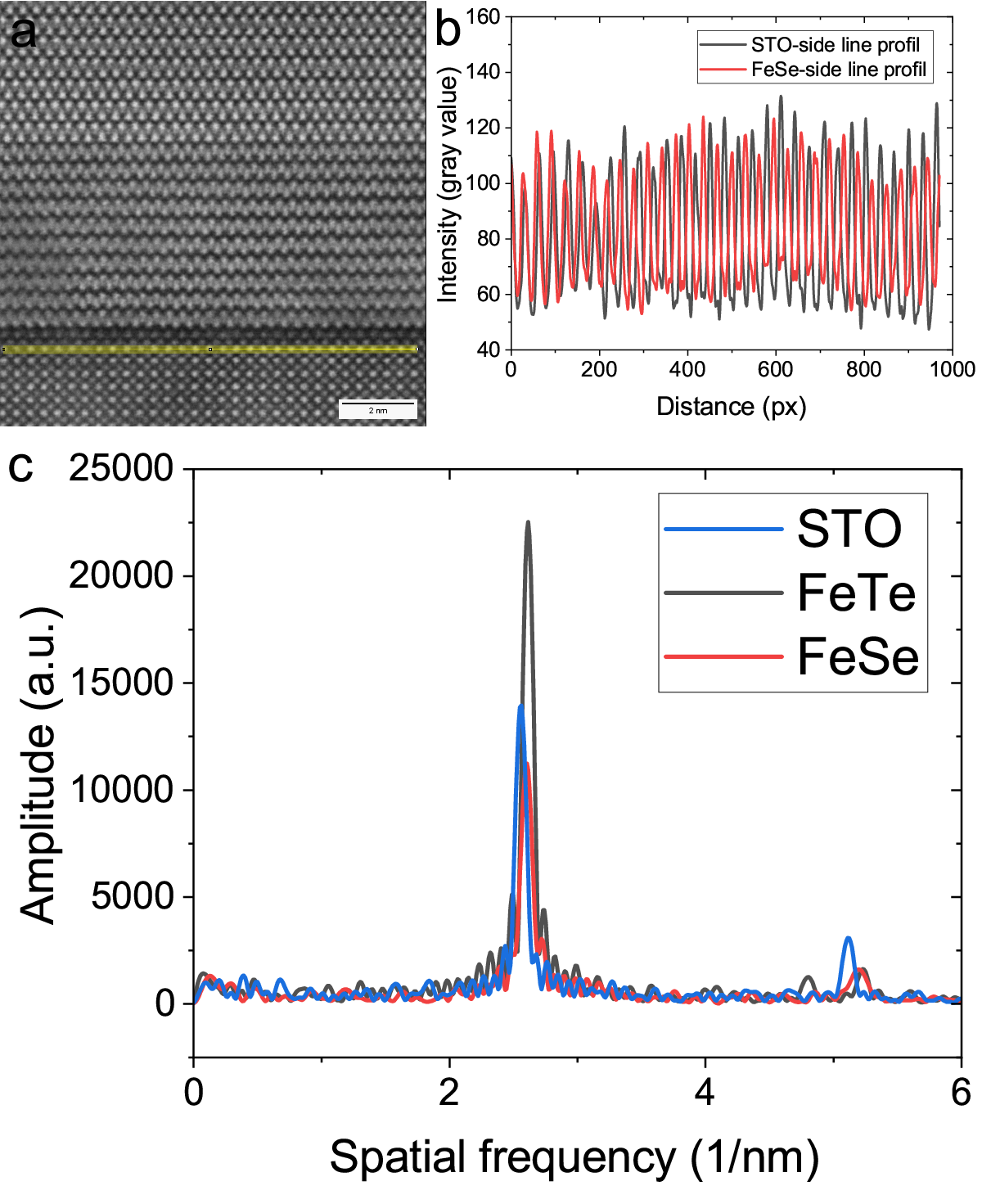}
  \caption{(a) HAADF-STEM image of the FeSe/SrTiO\textsubscript{3} interface with a horizontal trace (yellow) marking the atomic row along which the line profile was extracted. (b) Intensity profiles taken at the same lateral position on both sides of the STO--FeSe interface: the STO profile corresponds to the uppermost SrO layer, while the FeSe profile is taken on the adjacent first FeSe layer. The two traces exhibit an initially matched periodicity that gradually develops a lateral phase shift over approximately 60 unit cells. (c) One-dimensional Fourier amplitude spectra of three line profiles extracted directly at the three relevant interfaces: STO at the STO--FeSe interface, FeSe at the FeSe--STO interface, and FeTe at the FeTe--FeSe interface (taken on the lowest FeTe layer). The three spectra show clearly separated first-harmonic peak positions. The STO peak was fixed to 3.905 \AA for calibration, enabling direct extraction of the in-plane lattice constants of FeSe and FeTe.}
  \label{fig:S8}
\end{figure}

\end{document}